\documentstyle[12pt]{article}

\catcode`\@=11
\def\@citex[#1]#2{%
\if@filesw \immediate \write \@auxout {\string \citation {#2}}\fi
\@tempcntb\m@ne \let\@h@ld\relax \def\@citea{}%
\@cite{%
  \@for \@citeb:=#2\do {%
    \@ifundefined {b@\@citeb}%
      {\@h@ld\@citea\@tempcntb\m@ne{\bf ?}%
      \@warning {Citation `\@citeb ' on page \thepage \space undefined}}%
      {\@tempcnta\@tempcntb \advance\@tempcnta\@ne%
      \@tempcntb\number\csname b@\@citeb \endcsname \relax%
      \ifnum\@tempcnta=\@tempcntb 
        \ifx\@h@ld\relax%
          \edef \@h@ld{\@citea\csname b@\@citeb\endcsname}%
        \else%
          \edef\@h@ld{\ifmmode{-}\else--\fi\csname b@\@citeb\endcsname}%
        \fi%
      \else
        \@h@ld\@citea\csname b@\@citeb \endcsname%
        \let\@h@ld\relax%
      \fi}%
    \def\@citea{,\penalty\@highpenalty\,}%
  }\@h@ld
}{#1}}
\newcount\hour
\newcount\minute
\newtoks\amorpm
\hour=\time\divide\hour by 60
\minute=\time{\multiply\hour by 60 \global\advance\minute by-\hour}
\edef\standardtime{{\ifnum\hour<12 \global\amorpm={am}%
	\else\global\amorpm={pm}\advance\hour by-12 \fi
	\ifnum\hour=0 \hour=12 \fi
	\number\hour:\ifnum\minute<10 0\fi\number\minute\the\amorpm}}
\edef\militarytime{\number\hour:\ifnum\minute<10 0\fi\number\minute}

\def\draftlabel#1{{\@bsphack\if@filesw {\let\thepage\relax
   \xdef\@gtempa{\write\@auxout{\string
      \newlabel{#1}{{\@currentlabel}{\thepage}}}}}\@gtempa
   \if@nobreak \ifvmode\nobreak\fi\fi\fi\@esphack}
	\gdef\@eqnlabel{#1}}
\def\@eqnlabel{}
\def\@vacuum{}
\def\marginnote#1{}
\def\draftmarginnote#1{\marginpar{\raggedright\scriptsize\tt#1}}

\def\draft{\oddsidemargin -.5truein
	\def\@oddfoot{\sl DRAFT \hfil
	\rm\thepage\hfil\sl\today\quad\militarytime}
	\let\@evenfoot\@oddfoot \overfullrule 3pt
	\let\label=\draftlabel
	\let\marginnote=\draftmarginnote
   \def\@eqnnum{(\theequation)\rlap{\kern\marginparsep\tt\@eqnlabel}%
\global\let\@eqnlabel\@vacuum}  }

\@addtoreset{equation}{section}
\@addtoreset{footnote}{section}
\def\theequation{\thesection.\arabic{equation}}
\def\section{\@startsection {section}{1}{\z@}{3.ex plus 1ex minus
 .2ex}{2.ex plus .2ex}{\large\bf}}
\def\subsection{\@startsection{subsection}{2}{\z@}{2.75ex plus 1ex minus
 .2ex}{1.5ex plus .2ex}{\bf}}

\def\appendix{\let\appendix\moreappendix%
	\setcounter{section}{0} \setcounter{subsection}{0}
	\@addtoreset{theorem}{section}
	\def\thesection{\Alph{section}}\moreappendix}
\def\moreappendix{\@startsection {section}{1}{\z@}{3.ex plus 1ex minus
 .2ex}{2.ex plus .2ex}{\large\bf Appendix }}

\def\eqalign#1{\null\,\vcenter{\openup\jot\m@th
  \ialign{\strut\hfil$\displaystyle{##}$&$\displaystyle{{}##}$\hfil
      \crcr#1\crcr}}\,}
\def\eqalignno#1{\displ@y \tabskip\centering
  \halign to\displaywidth{\hfil$\@lign\displaystyle{##}$\tabskip\z@skip
    &$\@lign\displaystyle{{}##}$\hfil\tabskip\centering
    &\llap{$\@lign##$}\tabskip\z@skip\crcr
    #1\crcr}}

\catcode`\@=12

\def\eq#1{.~(\ref{#1})}
\def\beq{\begin{equation}}
\def\eeq{\end{equation}}
\def\beqar{\begin{eqnarray}}
\def\eeqar{\end{eqnarray}}

\def\nfrac#1#2{{\displaystyle{\vphantom1\smash{\lower.5ex\hbox{\small$#1$}}%
	\over\vphantom1\smash{\raise.25ex\hbox{\small$#2$}}}}}

\def\p#1{\mskip#1mu}
\def\n#1{\mskip-#1mu}
\def\stop{\p6.}
\def\comma{\p6,}
\def\semi{\p6;}
\def\eqand{\p6 {\rm and}}

\def\NN{N \n{14} N}
\def\ZZ{Z \n{10} Z}

\def\noj#1,#2,{{\bf #1} (19#2)\ }
\def\jou#1,#2,#3,{{\sl #1\/ }{\bf #2} (19#3)\ }
\def\ann#1,#2,{{\sl Ann.\ Physics\/ }{\bf #1} (19#2)\ }
\def\cmp#1,#2,{{\sl Comm.\ Math.\ Phys.\/ }{\bf #1} (19#2)\ }
\def\cq#1,#2,{{\sl Class.\ Quantum Grav.\/ }{\bf #1} (19#2)\ }
\def\cqg#1,#2,{{\sl Class.\ Quantum Grav.\/ }{\bf #1} (19#2)\ }
\def\ijmp#1,#2,{{\sl Int.\ J.\ Mod.\ Phys.\/ }{\bf A#1} (19#2)\ }
\def\jmp#1,#2,{{\sl J.\ Math.\ Phys.\/ }{\bf #1} (19#2)\ }
\def\grg#1,#2,{{\sl Gen.\ Rel.\ Grav.\/ }{\bf #1} (19#2)\ }
\def\mpl#1,#2,{{\sl Mod.\ Phys.\ Lett.\/ }{\bf A#1} (19#2)\ }
\def\nc#1,#2,{{\sl Nuovo Cim.\/ }{\bf #1} (19#2)\ }
\def\np#1,#2,{{\sl Nucl.\ Phys.\/ }{\bf B#1} (19#2)\ }
\def\pl#1,#2,{{\sl Phys.\ Lett.\/ }{\bf #1B} (19#2)\ }
\def\pla#1,#2,{{\sl Phys.\ Lett.\/ }{\bf #1A} (19#2)\ }
\def\pr#1,#2,{{\sl Phys.\ Rev.\/ }{\bf #1} (19#2)\ }
\def\prd#1,#2,{{\sl Phys.\ Rev.\/ }{\bf D#1} (19#2)\ }
\def\prl#1,#2,{{\sl Phys.\ Rev.\ Lett.\/ }{\bf #1} (19#2)\ }
\def\prp#1,#2,{{\sl Phys.\ Rept.\/ }{\bf #1C} (19#2)\ }
\def\ptp#1,#2,{{\sl Prog.\ Theor.\ Phys.\/ }{\bf #1} (19#2)\ }
\def\ptpsup#1,#2,{{\sl Prog.\ Theor.\ Phys.\/ Suppl.\/ }{\bf #1} (19#2)\ }
\def\rmp#1,#2,{{\sl Rev.\ Mod.\ Phys.\/ }{\bf #1} (19#2)\ }
\def\yadfiz#1,#2,#3[#4,#5]{{\sl Yad.\ Fiz.\/ }{\bf #1} (19#2) #3%
\ [{\sl Sov.\ J.\ Nucl.\ Phys.\/ }{\bf #4} (19#2) #5]}
\def\zh#1,#2,#3[#4,#5]{{\sl Zh.\ Exp.\ Theor.\ Fiz.\/ }{\bf #1} (19#2) #3%
\ [{\sl Sov.\ Phys.\ JETP\/ }{\bf #4} (19#2) #5]}
\def\ie{\hbox{\it i.e.\/}}

\def\etal{{\it et al.\/}}

\def\apriori{{\it a priori\/}}

\def\ket#1{\left| #1 \right \rangle}

\def\upda{\updownarrow}
\def\da{\downarrow}
\def\ua{\uparrow}
\def\f{\frac}
\def\pa{\partial}

\def\to{\rightarrow}

\newtheorem{lemma}{Lemma}
\newtheorem{theorem}{Theorem}
\newtheorem{qtheorem}[theorem]{Claim}

\def\to{\rightarrow}
\def\To{\longrightarrow}
\def\longlongrightarrow{\relbar\joinrel\relbar\joinrel\rightarrow}
\def\ridiculousrightarrow{\relbar\joinrel\relbar\joinrel\relbar%
\joinrel\rightarrow}

\def\onnearrow#1{\mathrel{\mathop{\nearrow}\limits^{#1}}}
\def\undernearrow#1{\mathrel{\mathop{\nearrow}\limits_{#1}}}
\def\onarrow#1{\mathrel{\mathop{\longrightarrow}\limits^{#1}}}
\def\onArrow#1{\mathrel{\mathop{\longlongrightarrow}\limits^{#1}}}
\def\OnArrow#1{\mathrel{\mathop{\ridiculousrightarrow}\limits^{#1}}}

\catcode`\@=11
\def\l:{\mathopen{:}\,}
\def\r:{\,\mathclose{:}}
\@addtoreset{footnote}{subsection}
\def\heading#1{\par\penalty-1000{\noindent \bf #1}}
\catcode`\@=12

\def\calp{{\cal P}}
\def\calx{{\cal X}}
\def\calo{{\cal O}}
\def\gh{{\zeta\eta}}
\def\upgh{\otimes\ket{\uparrow}_\gh}
\def\downgh{\otimes\ket{\downarrow}_\gh}
\def\tb{\tilde{b}}
\def\teta{\tilde\eta}
\def\half{\hbox{$\f12$}}
\def\thrhalf{\hbox{$\f32$}}
\def\fhalf{\hbox{$\f52$}}
\def\spin{\hbox{spin}}

\begin{document}
\begin{titlepage}

\begin{center}
June 3, 1992 \hfill    TAUP--1962--92 \\
\hfill    hep-th@xxx/9206014

\vskip 1 cm

{\large \bf
Discrete States of 2D String Theory in Polyakov's Light-Cone Gauge
}

\vskip 1 cm

\vskip 1.2cm
{
          Neil Marcus {\it and\/} Yaron Oz\footnote{
Work supported in part by the US-Israel Binational Science Foundation,
and the Israel Academy of Science.\\E-Mail:
NEIL@HALO.TAU.AC.IL, YARONOZ@TAUNIVM.BITNET.}
}
\vskip 0.2cm

{\sl
School of Physics and Astronomy\\Raymond and Beverly Sackler Faculty
of Exact Sciences\\Tel-Aviv University\\Ramat Aviv, Tel-Aviv 69978, ISRAEL.
}

\end{center}

\vskip 1 cm

\begin{abstract}

We find the discrete states of the $c=1$ string in the light-cone gauge of
Polyakov.  When the state space of the gravitational sector of the theory
is taken to be the irreducible representations of the $SL(2,R)$ current
algebra, the cohomology of the theory is {\it not\/} the same as that in the
conformal gauge.  In particular, states with ghost numbers up to 4
appear.  However, after taking the space of the theory to be the
Fock space of the Wakimoto free-field representation of  the $SL(2,R)$, the
light-cone and conformal gauges are equivalent.  This supports the
contention that the discrete states of the theory are physical. We point out
that the natural states in the theory do not satisfy the KPZ constraints.

\end{abstract}
\newpage

\end{titlepage}

\section{Introduction}

Much effort has been devoted in recent years to the study of conformal
matter systems coupled to two-dimensional quantum gravity. The barrier at
$c=1$, first discovered by Knizhnik, Polyakov and Zamolodchikov in the
light-cone gauge \cite{Zam} and confirmed later by David, Distler and Kawai
in the conformal gauge \cite{David,Distler}, has restricted the study mainly
to couplings to matter systems with central charge $c\leq 1$. Nevertheless,
it turned out that these theories possess rich and quiet non-trivial
structure and symmetries.

Among these models the $c=1$ case is exceptional in that it has a
two-dimensional space-time interpretation.  In addition to the tachyon, the
spectrum of this model consists of infinite set of discrete states, present
for quantized values of momentum \cite{Lian,Pilch}. These special states
appeared in the calculation of puncture operators correlation functions by
Gross, Klebanov and Newman \cite{Gross}, and were found in the continuum by
Polyakov, who interpreted them as remnants of transverse string excitations
\cite{resonant}.  Witten showed that the spin zero ghost number zero discrete
states generate a ground ring, and that one can combine antiholomorphic
spin zero and holomorphic spin one states to yield $w_{\infty}$ symmetry
currents of the theory \cite{Witten}.  These currents have been used to
facilitate the calculation of tachyon amplitudes \cite{ward}. A $w_\infty$
symmetry structure has also been uncovered in the $c=1$ matrix model
\cite{Moore,MPY,Jev,Das}.

Yet, it is not clear whether the discrete states are indeed physical
objects.  Although they appear as poles in tachyon amplitudes
\cite{Gross,resonant,diFrancesco,minic}, one can interpret these poles
simply as a renormalization of the external legs.  Also, the correlation
functions, in the continuum, of the discrete states themselves blow-up, and
seem to vanish upon regularization \cite{Li}.

Evidently, if discrete states are physical they should appear in any
legitimate gauge fixing. So far, in the continuum, they have been found and
analyzed in the conformal gauge, with the Liouville field treated as a
free scalar with a background charge\cite{Lian,Pilch,Witten,Kleb,Zwie,Sen}.
The aim of this paper is to study the spectrum of the theory in Polyakov's
light-cone gauge.  Previous works in the literature aiming at the analysis
of the spectrum of conformal matter coupled to gravity theories in the
light-cone gauge exist \cite{Itoh}, but do not reveal the structure of the
discrete states in the spectrum. We will be interested in finding the
discrete states and analyzing their structure in this gauge, and comparing
the results to the conformal gauge.  As we will discuss at the end of the
paper, our results are applicable for all $c \le 1$. However, we shall mainly
deal with the $c=1$ case.

The paper is organized as follows: In section~2 we state the BRST cohomology
problem in the light-cone gauge and set the notations and conventions. In
section~3 we start analyzing the BRST cohomology, taking the state space of
the gravitational sector to be the irreducible Kac-Moody module of
Polyakov's residual $SL(2,R)$ light-cone gauge current algebra. We find the
vacuum and tachyon states, and see that the states in the light-cone gauge
fall into pairs of conjugate states.  We also note
that the KPZ constraints do not hold for the sector built on the vacuum.   In
section~4 we analyze the cohomology using a free field representation of the
$SL(2,R)$, taking the state space to be the Fock space of the free
fields.  We again find a cohomology module and a module dual to it, and
we prove that the cohomology is equivalent to that of the conformal gauge.
Finally, we  calculate the light-cone analogues of the generators of
Witten's ground ring, as well as the other operators in the cohomology.   In
section~5 we return to the cohomology on the Kac-Moody module.  We examine the
first level explicitly, and see that extra states appear in the
current algebra cohomology that are not in the cohomology of the Fock
space.  We then
develop the ``Felder resolution'', that allows us to obtain
the complete current algebra cohomology from that of the Fock space.
We find that there are discrete states in the Kac-Moody case
with ghost numbers ranging from $-2$ to $4$.  This approach
is therefore {\it not\/} equivalent to the
conformal gauge.  Section~6 is devoted to discussion and conclusions.

\section{The BRST cohomology problem in the light-cone gauge.}

An action describing the $c=1$ conformal field theory coupled to
two-dimensional gravity is given by:
\beq
S = \int{\rm d}^2z\sqrt{g}g^{ab}{\pa}_ax{\pa}_bx \comma  \label{action}
\eeq
where $g^{ab}$ is the two-dimensional metric and $x$ is a scalar field.

In the conformal gauge, the metric is fixed to the form $g_{ab} = e^\phi
\delta_{ab}$, where $\phi$ is the Liouville field.  The vanishing of the Weyl
anomaly leads to the equation
\beq
\nabla^2 \phi=0 \comma
\eeq
where we have set the cosmological constant to zero .  The main problem with
the conformal gauge is that, while $\phi$ appears like an ordinary massless
scalar field, its quantization is not straightforward, since its measure in the
functional integral is field dependent. Nevertheless, the quantization has
been carried out with great success using the David, Disler and Kawai {\it
ansatz\/} of considering $\phi$ to be a free scalar with a background charge
fixed at the quantum level \cite{David,Distler}. In this gauge the fields of
the theory, in addition to $\phi$, are the matter field $x$, and the
standard $(b,c)$ and $(\bar{b},\bar{c})$ ghost systems associated with the
secondary constraints $T_{zz}=T_{\bar{z}\bar{z}}=0$, where $T_{ab}$ is the
energy-momentum tensor.

In Polyakov's light-cone gauge the metric is gauge-fixed to
\beq
ds^2 = d z d{\bar z} + h(z,\bar{z}) d z d z \stop
\eeq
The nature of the metric implies that the measure of $h$ is not field
dependent, so one does not encounter the difficulties of the Liouville
field.  In this case the gauge fixing conditions $g_{{\bar z}{\bar z}} = 0$
and $g_{{\bar z}z} = \half $ lead to the secondary constraints:
\beq
T_{zz} = 0 \qquad \eqand \qquad T_{z{\bar z}} = 0 \stop \label{constraints}
\eeq
We denote the ghost systems associated with these constraints by $(b,c)$
and $(\zeta,\eta)$,  respectively.  They are both anticommuting, and
have spins $(2,-1)$ and $(0,1)$.

The vanishing of the gravitational anomaly leads to the equation
\beq
\pa_{\bar{z}}^3 h(z,\bar{z}) = 0 \comma
\eeq
so $h$ is decomposed into three parts:
\beq
h(z,{\bar z}) = J^+(z) + 2{\bar z}J^0(z) + ({\bar z})^2J^-(z) \stop
\label{decomp}
\eeq
As was shown by Polyakov, by analyzing the Ward identities of the theory,
the $J$'s satisfy an $SL(2,R)$ Kac-Moody algebra \cite{Pol}, with OPE's:
\beq
J^a(z)J^b(w) \sim \f{-\f\kappa2 \, \eta^{a b}}{(z-w)^2} + \f{f^{ab}_c}
{z-w}J^c(w) \stop \label{sl2r}
\eeq
Here $\kappa$ is the (renormalized) Kac-Moody central charge, the
$f^{ab}_c$'s are the structure constants of $SL(2,R)$ algebra, and
$\eta^{ab}$ its Cartan-Killing form\footnote{In our notation, $f^{+-}_0 =
2, f^{0+}_+ = -1, f^{0-}_- = 1$ and $\eta^{+-}=2, \eta^{00}=-1$.}.
The operator product expansion of the other fields of the theory are given by:
\beqar
x(z)x(w) &\sim& -\log (z-w) \nonumber \\
b(z)c(w) &\sim &\f1{z-w} \nonumber \\
\zeta(z)\eta(w) &\sim& \f1{z-w} \stop
\eeqar

The ``holomorphic'' part of the energy-momentum tensor of the theory
$T(z) \, \equiv \, T_{zz}(z)$ is given by
\beq
T(z) = T^{grav}(z) + T^{matt}(z) + T^{bc}(z) +
T^{\gh}(z) \comma
\eeq
where $T^{grav}(z), T^{matt}(z), T^{bc}(z)$ and $T^{\gh}(z)$ are the
stress tensors of the gravity, matter and ghost sectors respectively. The
stress tensors of the matter and ghost systems are, as usual, given
by\footnote{Throughout the paper, we use conformal field theory normal
ordering.}:
\beqar
& T^{matt}(z) &= -\half  \, \l: ({\pa}x)^2 \r: \nonumber\\
& T^{bc}(z) &= \l: -2b{\pa}c -{\pa}bc \r: \nonumber\\
& T^{\gh}(z) &= \l: {\pa}\gh \r: \stop
\eeqar
Knizhnik, Polyakov and Zamolodchikov (KPZ) showed that the gravity stress
tensor takes the form of a modified Sugawara construction \cite{Zam}:
\beq
T^{grav}(z) = -\f1{\kappa+2}\, \eta_{ab} \,
               \l: J^a(z)J^b(z) \r: ~+~ {\pa}J^0(z) \comma
\label{sugawara}
\eeq
where the second piece modifies the spins of the $SL(2,R)$ Kac-Moody
currents such that they are compatible with the decomposition
(\ref{decomp}). Thus $\spin(J^+)=2$, $\spin(J^-)=0$ and $J^0$ remains of
spin 1, but is no longer a primary field:
\beq
T(z) J^0 (w) \sim -\f\kappa{(z-w)^3} + \f{J^0(w)}{(z-w)^2} +
\f{\pa J^0(w)}{z-w}         \label{not primary} \stop
\eeq

The component of the gravity stress tensor $T_{z{\bar z}}$ is
essentially the Kac-Moody current $J^-$ \cite{Zam}:
\beq
T_{z{\bar z}} \sim \pa_{\bar z}^2 h
\sim J^-(z)  \stop \label{tzzbar}
\eeq
Thus the constraints algebra---the algebra of the residual symmetry of the
light-cone gauge---takes the form:
\beqar
T(z)T(w) &\sim& \f{c/2}{(z-w)^4} + \f{2T(w)}{(z-w)^2} +
             \f{\pa{T(w)}}{z-w} \nonumber\\
T(z)J^-(w) &\sim& \f{\pa{J^-(w)}}{z-w} \nonumber\\
J^-(z)J^-(w) & \sim & 0 \vphantom{\f{\pa{J^-(w)}}{z-w}} \stop \label{constr}
\eeqar
The contribution to the central charge from the gravity sector, $c^{grav}$,
can be found using eq\eq{sugawara} to be $c_{grav}=
3\kappa/(\kappa+2)-6\kappa$; $c^{matt} = 1$, $c^{bc} = -26$  and $c^\gh =
-2$.  The total central charge vanishes if
\beq
\kappa=-3 \stop
\eeq
The system of constraints is then of first class type, and it is natural to
impose them via the BRST formalism. The BRST operator corresponding to the
algebra of constraints (\ref{constr}) is:
\beq
Q_B = \int\f{{\rm d}z}{2{\pi}i} \, \l: c(z) \left(T^{grav}(z) +T^{matt}(z) +
\half T^{bc}(z)
+ T^{\gh}(z) \right ) + \eta(z) J^-(z) \r: \label{BRST}
\eeq
The spectrum of the theory is then given by the cohomology of the BRST
complex $(Q_B,\cal{H})$ where the quantum state space ${\cal H}$ is
decomposed into ${\cal H}_{grav} \otimes{\cal H}_{matt} \otimes{\cal H}_{bc}
\otimes{\cal H}_ {\gh}$.  The matter and ghost state spaces are
simply the Fock spaces built out of the oscillator modes of the fields.
However, it is not \apriori{} apparent what should be chosen
for the gravity sector state space.

\section{BRST cohomology on the $SL(2,R)$ Kac-Moody module}
\subsection{Zero modes of the $SL(2,R)$.}

The perhaps most natural choice for the gravity sector state space
${\cal H}_{grav}$
is to take it to be the irreducible $SL(2,R)$ Kac-Moody modules
built out of the Kac-Moody currents.   States in a Kac-Moody module can be
described by acting creation modes of the currents, $J_{-n}^a$, on a
member of a representation of the $SL(2,R)$ Lie algebra generated by the
zero-modes $J_0^a$.  Because the Sugawara stress tensor
of eq\eq{sugawara} is twisted,
the central term of the $SL(2,R)$ Kac-Moody algebra is modified:
\beq
[J_n^a,J_m^b] = f^{ab}_cJ^c_{n+m} -\f{\kappa}2\eta^{ab}(n+a)\delta_{n+m,0}
\comma \label{comm}
\eeq
where ``$n+a$'' denotes $n \pm 1$ for $a=\pm$, and $n$ for $a=0$.  The
$J_0^a$'s thus do not quite generate $SL(2,R)$.  It is therefore
convenient to define new generators\footnote{Since there is no notion of
a unitary representation in this theory---in fact the states of the theory
do not fall into representations of $SL(2,R)$ at all---one may as well Wick
rotate the $SL(2,R)$ into the form of the more familiar $SU(2)$.  It should
be noted, however, that the hermiticity properties of the $j$'s are very
nonstandard.}
\beqar
j^\pm  &\equiv& i \, J_0^\mp \nonumber\\
j^z &\equiv& J_0^0 + \thrhalf \comma \label{su2}
\eeqar
with $SU(2)$-like commutation relations:
\beqar
[j^+,j^-] &=& 2 j^z \nonumber\\
{} 
[j_z,j^\pm] &=& \pm j^{\pm} \stop
\eeqar
Then, defining the usual $SU(2)$ Casimir operator
\beq
C \equiv J(J+1) =  ( j^{z} )^2 +\half \{ j^+,j^- \} \comma \label{Casi}
\eeq
one can represent the ground states of the gravity sector as $\ket{J,M}$ with
$M$ being the eigenvalue of $j^z$.

The ground states in the other sectors of the theory describe the value of
the $x$-momentum, $p$, and the usual $\ket{\upda}_{bc}$ of the
zero-modes of the $(b,c)$ system.  One also needs to define
$\ket{\upda}_\gh$ for the $(\zeta,\eta)$ ghosts:
\beqar
\zeta_0 \ket{\da}_{\gh} &=& 0 \nonumber\\
\eta_0 \ket{\ua}_{\gh} &=& 0 \stop
\eeqar

\subsection{The KPZ condition}

Using the mode expansion of the fields, the BRST operator of eq\eq{BRST}
takes the form
\beq
Q_B = \sum_m \l: c_{-m} \left( L_m^{grav} +L_m^{matt} + \half L_m^{bc} +
L_m^{\gh} \right) + \eta_{-m} J^-_m \r: \comma
\eeq
where the $L_m$'s are the Virasoro generators of the various
sectors.  $L_n^{grav}$ is given by
\beq
L_n^{grav} = \f{-1}{\kappa+2} \sum_m \l: \left( \half J^-_m J^+_{n-m}+
\half J^+_m J^-_{n-m} - J^0_m J^0_{n-m} \right) \r:
- (n+1) \, J_n^0 \stop    \label{Ln}
\eeq
The other $L_n$'s are given explicitly in appendix~A. As usual, the first
step in finding the cohomology of $Q_B$ is to decompose it with respect to the
$b,c$ zero modes:
\beq
Q_B = c_0 L_0 - b_0 \sum_{n\neq 0} n \, c_{-n} c_n + \hat{Q} \stop
\eeq
Since $L_0 = \{Q_B,b_0\}$, the cohomology of $Q_B$ is contained in
the kernel of $L_0$ \cite{bosonic}\footnote{If $L_0 \Psi \neq 0$, and $\Psi$ is
closed, then $\Psi$ is also exact: $\Psi= Q_B (L_0)^{-1} b_0 \Psi$.}.
This means that one can
first calculate the relative cohomology, \ie{} the cohomology of $\hat{Q}$
restricted to the subspace of states $\ket\psi$ satisfying
$b_0\ket\psi = 0$,  and then use this to derive the absolute cohomology.

$L_0$ can be found explicitly by carefully performing the normal ordering
in the definitions of the $L$'s, and using the definition of
the Casimir operator $C$ in eq\eq{Casi}:
\beq
L_0 = \half  \, p^2 - C -\f1{4} + N \comma
\eeq
where $N$ is the level operator measured from the tachyon ground state.  The
appearance of the Casimir operator is not surprising, since $L_0$ commutes
with the generators of the $SL(2,R)$. The requirement that $L_0 = 0$ can be
rewritten as
\beq
\half  \, p^2 + N = (J + \half )^2 \stop \label{L0}
\eeq
This can be compared to the KPZ relation \cite{Zam}:
\beqar
\Delta_0 &=& \Delta + \f{\Delta(1-\Delta)}{\kappa+2} \nonumber\\
\onArrow{\kappa=-3} {\Delta}_0 &=& \Delta^2 \stop \label{KPZ}
\eeqar
Here ${\Delta}_0$ is the dimension of the undressed physical operator, and
$\Delta$ is its dimension dressed by the gravitational fluctuations.  The
KPZ formula is derived for states that satisfy the constraints
(\ref{constraints}) of the theory.  Since $T_{z{\bar z}} \sim J^-(z)$,
this means that such states
are annihilated by $J_0^- \sim j^+$, and are therefore heighest weight
states (HWS).   Comparing eqs\eq{L0} and (\ref{KPZ}), and noting that HWS
satisfy $J=M=J_0^0+\thrhalf $, one sees that
\beq
\Delta = J_0^0+2 \stop
\eeq
The reason for this shift, which is not apparent in the original paper of
Polyakov \cite{Pol}, is that one needs to measure the dressed
dimensions of operators with respect to that of the cosmological
constant operator.  A similar shift appears in the conformal gauge
\cite{Distler}.  The (HWS) cosmological constant operator, a tachyon with
$p=0$, has $J=-\half$ and therefore $J_0^0=-2$.

\subsection{The $SL(2,R)$ invariant vacuum}

Before continuing with the general problem of finding the cohomology, it is
instructive to consider the (conformal field theory, not Kac-Moody!)
$SL(2,R)$ invariant vacuum of the theory, $\ket0$, defined by $A(0) \ket0
\sim \hbox{regular}$ for all primary fields $A(z)$.
As one would expect, the vacuum is indeed in the cohomology of
$\hat{Q}$.  This is shown in appendix~A.  Examining the spins of the fields in
the theory, one sees that $\ket0$ is annihilated by $p$, $b_0$, $b_{-1}$,
$\eta_0$, $J_0^0 = j^z-\thrhalf$, $J_0^+ = j^-$ and
$J_{-1}^+$\footnote{Actually,
as can be seen by eq\eq{not primary}, $J^0$ is not a primary field.
However $J_0^0$ still annihilates $\ket0$, since both $J_{-1}^+$ and
$J_1^-$ annihilate it, and $J_0^0$ is proportional to their commutator.},
and can therefore be written as
\beq
\ket0 =  b_{-1}\ket{ p=0, J=-\thrhalf  , M=\thrhalf  } \otimes \ket{\ua}_\gh
\otimes \ket{\da}_{bc}  \stop
\eeq
The vacuum therefore has several peculiar properties: First, it carries
$M=3/2$, so the theory has background charge 3.  Second, since its $J$ is
negative, it is not an element of a finite representation of $SL(2,R)$ but
of a semi-infinite one. Third,  since it is annihilated by $j^-$ it is a
{\it lowest\/} weight state, with $J=-M$.    {\it The vacuum
of the theory is not a KPZ state!\/}

\subsection{Zero'th-level states (tachyons)}

At the lowest level of the theory states have no oscillators.  The
relative cohomology operator $\hat{Q}$ therefore reduces to
\beq
\hat{Q} \sim -i\,  \eta_0 \, j^+ \stop
\eeq
In the relative cohomology there are only two possible types of
states: $\ket{p,J,M}\downgh$ and $\ket{p,J,M}\upgh$.
(The $\ket{\da}_{bc}$ state is not explicitly written.)   Now
\beqar
\hat{Q} \left( \ket{p,J,M} \downgh \right) &=&
                 -i \,  j^+ \ket{p,J,M} \upgh    \nonumber\\
\hat{Q} \left( \ket{p,J,M} \upgh  \right) &=& 0 \stop
\eeqar
The first state is therefore closed if it is a heighest-weight state (HWS),
with $M=J$, while the second state is automatically closed.  Also, while
the first state can never be exact, the second state is exact unless it
cannot be written as $j^+ \ket{p,J,M}$. This means that it is a
lowest-weight state (LWS), with $M=-J$.
The value of $J$ is determined from the $L_0=0$ condition
(eq\eq{L0}) with $N=0$ to be either of the two solutions
\beq
J^\pm(p) = \pm \f{|p|}{\sqrt2}- \half  \comma \label{J}
\eeq
and the relative cohomology is represented by the states
\beqar
&&\ket{p,J^\pm(p) ,\phantom{-}J^\pm(p) } \otimes \ket\da_\gh \nonumber\\
&&\ket{p,J^\pm(p) ,-J^\pm(p) } \otimes \ket\ua_\gh \stop
\eeqar
These states correspond to the tachyon, since they exist for all $p$.

This result can be compared to the tachyons of the conformal gauge, which
have the form\footnote{We use the radially-ordered notation that the vacuum has
$p_\phi=0$, rather than Witten's symmetric notation that it carries $p_\phi=i
\sqrt2$.}
\beq
\ket{p, p_\phi^\pm= -i ( \sqrt2 \pm |p| ) } \stop
\eeq
We see that, because of the zero modes of the  $(\zeta,\eta)$ ghost system,
there appears to be a doubling of states in the light-cone gauge.  However,
it should be noted that, in general, the HWS are the largest members of
semi-infinite dimensional representations with $M \le J$,  while the LWS are
the smallest members of semi-infinite dimensional representations with $M
\ge J$.  These representations are conjugate, and can not be obtained from
each other.  (The only exception to this is when $2J+1 \in \NN$,
giving a finite dimensional representation with both a HWS
and a LWS.  This case corresponds to (half of) the ``discrete tachyons'' of the
conformal gauge.)  Since the vacuum is a LWS,
only states built on LWS can be obtained from it, and their
conjugate states should not be considered to give a duplication of the
spectrum of the theory.  Note that, as in the case of the vacuum state,  the
LWS representing the tachyons again do not satisfy the KPZ constraint
$J_0^- = 0$.

\heading{Absolute cohomology tachyons}

As is usual for tachyonic states, the absolute cohomology is obtained by simply
taking the relative cohomology states and the states obtained from them by
the raising operator $c_0$.  The general level-zero cohomology is therefore
given by the four types of states
\beqar
&&\ket{p,J^\pm(p) ,J^\pm(p) } \otimes \ket\da_\gh
                 \otimes\ket\upda_{bc}  \nonumber\\
&&\ket{p,J^\pm(p) ,-J^\pm(p) } \otimes \ket\ua_\gh
                 \otimes\ket\upda_{bc}   \stop
\eeqar

It is not simple to solve the cohomology at higher levels in a
straightforward way.  We shall therefore turn to the use of free field
representations of the current algebra in the next section, and return to
the cohomology of the current algebra in section 5.

\section{The BRST cohomology module in the Fock space}

\subsection{The Wakimoto free field representation}

Free field representations of conformal field theories are very useful for the
construction of operators and for the calculation of correlation functions
\cite{Feigin,Dotsenko}.  In this section we shall introduce the
Wakimoto representation for the $SL(2,R)$ current algebra of the gravity
sector, and find the cohomology of the BRST operator on the Fock space of the
free fields. In the next section, we shall use this cohomology to find the
cohomology on the irreducible Kac-Moody modules via the Felder resolution
\cite{BFelder}.  In this section we shall simply consider the Fock space to be
the state space of the theory.

In the Wakimoto representation of $SL(2,R)$ Kac-Moody algebra
the currents are given by \cite{Wakimoto}
\beqar
J^+ &=& \beta\gamma^2 + \f2{\alpha_+}\gamma{\pa}\varphi +
\kappa{\pa}\gamma\nonumber\\
J^0 &=& \beta\gamma + \f1{\alpha_+}{\pa}\varphi\nonumber\\
J^- &=& \beta \comma \label{wakimoto}
\eeqar
where $2/\alpha_+^2 = -(\kappa+2) = 1$.  Since the spins of the currents
in our case have been modified, the bosonic fields $\beta$ and $\gamma$
should be taken to have spins 0 and 1 respectively, the reverse of the
usual case.  $\varphi$ is a scalar field with background charge
$Q = -\f{i}2 (\alpha_++\f2{\alpha_+}) = -\sqrt2i$.  Using the OPE's:
\beq
\gamma(z)\beta(w) = \f1{z-w} \quad \eqand \quad {\varphi}(z){\varphi}(w) =
- \log (z-w) \comma
\eeq
one can show that the Wakimoto currents satisfy the $SL(2,R)$
current algebra of eq\eq{sl2r}.

With this representation, the Sugawara stress tensor of eq\eq{sugawara}
takes the simple form:
\beq
T^{grav} = \l: {\pa}\beta\gamma -\half  ({\pa}\varphi)^2 +
i Q{\pa}^2\varphi \r: \comma
\eeq
and the BRST operator of eq\eq{BRST} becomes
\beq
Q_B = \int\f{{\rm d}^2z}{2{\pi}i} \l: c(z) \left( - \half (\pa x)^2 - \half
(\pa\varphi)^2 + i Q {\pa}^2\varphi - b{\pa}c - \half {\pa}bc +
{\pa}\beta\gamma + {\pa}\gh \right) + \eta(z) \beta(z) \r: \stop
\eeq
{\it Note that the fields and stress tensor of the theory
in the Wakimoto representation are in a one to one correspondence with  those
of the conformal gauge\footnote{At least with the holomorphic sector part of
the conformal gauge.  The structure of the antiholomorphic sector of the
light-cone gauge is somewhat obscure.}, with $\varphi$ the analogue of the
Liouville field, except for the addition of the bosonic spin $(0,1)$ fields
$(\beta,\gamma)$ and the fermionic spin $(0,1)$ fields $(\zeta,\eta)$.\/}
This leads one to expect that these extra fields will conspire to cancel the
effects of each other, leaving behind the structure of the conformal gauge.
This
argument is supported by the fact that the $\eta(z)\beta(z)$ piece of the
BRST operator, coming from the constraint $J^-(z) = 0$, can be viewed as a
BRST operator generating the fermionic symmetry transformation
\beq
\delta\gamma = - \varepsilon\eta \qquad \delta\zeta = \varepsilon\beta \comma
\eeq
in the
topological field theory with action\footnote{We would like to thank
J.~Sonnenschein for this point.}
\beq
S = \int(\pa\beta\gamma +\pa\gh) \stop
\eeq
We shall see that this argument is essentially correct, but the
situation is somewhat more complicated and the extra fields do appear
in the operators representing the cohomology.

\subsection{The BRST cohomology: Zero modes}

As in the case of the current algebra, the first step in finding the
cohomology is to remove the zero-modes of the fields.  After achieving this,
the procedure becomes relatively simple, and follows that of the conformal
gauge.  Once again, the zero modes of $b$ and $c$ are separated
out by decomposing $Q_B$ into:
\beq
Q_B = c_0 L_0 - b_0 \sum_{n\neq 0} n \, c_{-n} c_n + \hat{Q} \stop
\eeq
One can then obtain the absolute cohomology of $Q_B$ from the relative
cohomology of $\hat{Q}$ acting on states annihilated by $b_0$ using the
result
\cite{Pilch,Zwie}:
\begin{theorem}
States in the absolute cohomology of $Q_B$ are of the form
$\Psi$ or $a_0 \Psi$, where $\Psi$ is in the relative cohomology of $\hat{Q}$.
As in the conformal gauge
\beq
a = [ Q , \varphi ] = c\pa \varphi + \sqrt2\pa c \comma
\eeq
so $a_0$, the zero mode of $a$, is essentially the BRST invariant
inverse of $b_0$.
\label{a}
\end{theorem}
The proof of this result follows exactly the proof in the conformal gauge
\cite{Zwie}.

In the light cone, one still has to deal with the zero-modes
of $(\zeta, \eta)$ and $(\beta,\gamma)$.  The next step in doing
this is to note that $Q_B$ does not contain $\zeta_0$, and to separate out
the $\eta_0$ part of $\hat{Q}$:
\beq
\hat{Q} \, \equiv \, \eta_0\calx + \hat{d} \comma
\eeq
with
\beq
\calx= \beta_0 + \sum_{n\neq 0}nc_n\zeta_{-n} \stop
\eeq
We shall later denote $\gamma_0$ by $\calp$, to remind
ourselves that $\calx$ and $\calp$ have commutation relations similar
to those of a
coordinate and a momentum: $[\calx,\calp]=1$.  The operator $\hat{d}$ is
still complicated, being given by
\beq
\hat{d} = \sum_{n\neq 0} \l: c_{-n}
\left(L_n^{matt} + L_n^{grav} + {L_n^\prime}^\gh \right)
+\sum_{n\neq 0} \eta_{-n} \beta_n
- \half \sum_{m,n \neq 0 \atop n+m\neq 0}(m-n) c_{-m}c_{-n}b_{n+m} \r:
\comma
\eeq
where the prime on ${L_n^\prime}^\gh$ denotes the exclusion of the piece of
$L_n^\gh$ containing $\eta_0$.  $L_n^{grav}$ is given by
\beq
L_n^{grav} = \sum_m \l: ( -m\beta_m \gamma_{n-m} + \half {\varphi}_m
{\varphi}_{n-m}) \r: - Q (n+1){\varphi}_n \semi
\eeq
the other $L_n$'s are given in appendix~A.

While the zero modes of $\eta$ and $\beta$ have been explicitly
extracted,
$\hat{d}$ still contains $\gamma_0$, via its dependence on $L_n^{grav}$.
This dependence can be removed by performing a Bogolubov
transformation from $(\beta_0, b_n,\eta_n)$ to $(\calx,\tb_n,\teta_n)$, with:
\beqar
\tb_n &\equiv& b_n +n\zeta_n\gamma_0 \nonumber\\
\teta_n &\equiv& \eta_n +nc_n\gamma_0 \stop \label{tilding}
\eeqar
The new set of oscillators
($\tb_n$,$c_n$,$\zeta_n$,$\teta_n$,$\beta_n$,$\gamma_n$) have the same
commutation relations as the original oscillators and, in addition, all
commute with $\calx$ and $\calp$.  The reason for performing the
transformation is that:

\begin{lemma}
$\hat{d}$ is independent of $\calx$ and $\calp$, when
written in terms of the ``tilded'' oscillators.
\end{lemma}

\noindent
The lemma is proven by seeing that $\hat{d}$ commutes with $\calx$ and $\calp$.
The result for $\hat{d}$ is given in eqs\eq{dhats}.

Because of the zero modes of the commuting $(\beta,\gamma)$ system, the Fock
space of the theory splits into two conjugate infinitely degenerate Hilbert
spaces which are not connected:  the first contains states with an arbitrary
number of $\gamma_0$'s acting on $\ket{\beta_0=0}$; the second contains
states with
$\beta_0$'s acting on  $\ket{\gamma_0=0}$.   After the  Bogolubov
transformation, states are built either on $\ket{\calx = 0}$ or on
$\ket{\calp=0}$.
Since $\calx = \{ \hat{Q} ,\zeta_0\}$, one would expect that the cohomology
of $\hat{Q}$ lies entirely within the kernel of $\calx$. In fact, the situation
is slightly more complicated, since one should obtain both such states and
their conjugates
in the $\ket{\calp=0}$ sector. Using the result that $\hat{d}$ is
independent of $\calx$ and $\calp$, one can reduce the calculation of the
cohomology of $\hat{Q}$ to that of the
relative cohomology of $\hat{d}$, acting on the Fock space without the zero
modes $\calx$, $\calp$, $\zeta_0$ and $\eta_0$, by the following theorem:
\begin{theorem}
States in the cohomology of $\hat{Q}$ are in one of the
two conjugate forms:
\beq
\Psi = \ket\psi \otimes \ket{\calx = 0}\otimes \ket{\da}_{\gh}
	\label{down}
\eeq
or
\beq
\Psi = \ket\psi \otimes \ket{{\calp}=0}\otimes\ket{\ua }_{\gh}
	\comma
\eeq
where $\ket\psi$ is in the cohomology of $\hat{d}$, acting  on the Fock
space of the tilded oscillators without zero modes.
\label{b}
\end{theorem}

\noindent
{\bf Proof}:  Consider first a $\Psi$ built on the state
$\ket{\calx=0}$:
\beq
\Psi \equiv \sum_{n=0}^{\infty} \ket{\psi_n} \otimes \calp^n \ket{\calx=0}
\comma
\eeq
where the $\ket{\psi_n}$'s are independent of $\calx$ and $\calp$, but are not
necessarily annihilated by $\zeta_0$ or $\eta_0$.
Requiring that $\Psi$ be closed under $\hat{Q}$ yields the equations:
\beq
\hat{d}\ket{\psi_n} + (n+1)\eta_0 \ket{\psi_{n+1}} = 0 \label{closed} \stop
\eeq
Using $\calx = \{ \hat{Q} ,\zeta_0\}$ and $\calx \sim \pa_\calp$,
one sees that
\beq
\Psi =  \zeta_0 \, \eta_0 \, \ket{\psi_0} \otimes \ket{\calx = 0}
+ \hat{Q} \left( \zeta_0 \sum_{n=0}^\infty \f1{n+1} \ket{\psi_n} \otimes
\calp^{n+1} \ket{\calx=0} \right) \stop
\eeq
$\Psi$ is therefore equivalent to a state $\Psi' = \zeta_0 \, \eta_0 \,
\ket{\psi_0} \otimes \ket{\calx = 0}$, up to an exact state.
Since $\zeta_0\eta_0$ is simply the projection operator to the
state $\ket{\da}_{\gh}$, $\Psi'$ has the desired form of
eq\eq{down}.  Now using eq\eq{closed} on $\Psi'$, one sees that
$\ket\psi$ is closed under $\hat{d}$, and it is clear that $\Psi_1' \equiv
\Psi_2'~$ {\it iff\/} they differ by a state exact under $\hat{d}$.  The
theorem is therefore established for the states built on $\ket{\calx=0}$.
One can use a similar proof for the states built on $\ket{\calp=0}$,
or simply argue that the conjugate of a state in the cohomology must
also be in the cohomology.

As far as the spectrum of the theory is concerned, one should not consider the
existence of cohomologically nontrivial states built on both $\ket{\calx=0}$
and $\ket{\calp=0}$ as a duplication of the states of the theory.  The fact
that conjugate states can be in disjoint Hilbert spaces is well known, and
occurs, for example, with the commuting ghosts of the superstring.  In fact, as
in the case of the superstring, these sectors are just two of an
infinite number of sectors.  \marginnote{Look!}
In the superstring,
one usually further ``bosonizes'' the ghosts \cite{FMS}, and is left
with a system with no remaining bosonic fields.
However, while such a bosonization of the $(\beta,\gamma)$ system  may
be useful for amplitude calculations, one can find the physical spectrum
of the theory
simply by restricting oneself to a particular sector. Since the vacuum of the
theory $\ket0 = b_{-1} \ket{\calp=0} \otimes \ket{\ua}_\gh \ket{\da}_{bc} $ is
in the $\ket{\calp=0}$ sector, we shall choose to work in this sector, which
has the  advantage that states in it can be represented by operators.

\subsection{The relative cohomology of $\hat{d}$}

We now need to compute the relative cohomology of $\hat{d}$.
This analysis will bear a close resemblance to the BRST analysis of the
$c=1$ theory in the conformal gauge by Bouwknegt, McCarthy and Pilch
\cite{Pilch}, and we shall follow their notations and proofs.  Introduce
the light-like combinations
\beq
\alpha_n^{\pm} = \f1{\sqrt2}(x_n\pm i{\varphi}_n) \qquad  n\neq 0 \comma
\eeq
where
\beqar
i\pa\varphi(z) &\equiv& \sum_n \varphi_n z^{-n-1} \nonumber\\
i\pa x(z)  &\equiv& \sum_n x_n z^{-n-1} \stop
\eeqar
The oscillators $\alpha_n^{\pm}$ satisfy the commutation relations
\beq
[\alpha_m^{\pm},\alpha_n^{\mp}] = m{\delta}_{n+m}.
\eeq
Define also
\beqar
p^{\pm} \phantom{(n)} &=& \f1{\sqrt2}(p\pm i(p_\varphi + i \sqrt2 ))\nonumber\\
P^{\pm}(n) &=& \vphantom{\f1{\sqrt2}} p^\pm \mp n \stop \label{pn}
\eeqar
Then $L_0$ becomes
\beqar
L_0 &=& p^+p^- + \sum_{n\neq 0} \l: \alpha_{-n}^+\alpha_n^- +n(c_{-n}\tb_n
       + \beta_{-n}\gamma_n + \zeta_{-n}\teta_n) \r: + 1
       \nonumber\\
    &\equiv& p^+p^- + \hat{L}_0 \comma
\eeqar
where $\hat{L}_0$ is the level operator for the oscillators
with respect to the Fock-space vacuum.

The next step is to impose a grading of the state space,
compatible with the algebra of the theory, so that
$\hat{d}$ breaks into a finite sum $\hat{d} = \sum_{i \ge 0} \hat{d}_i$,
where $\hat{d}_i$ has degree $i$.
One can then first compute the cohomology of $\hat{d}_0$ and then use a
basic result from the cohomology theory of ``filtered complexes''
to derive the cohomology of $\hat{d}$.  In order to obtain a simple
operator $\hat{d}_0$ we, extending the
results of the conformal gauge, define the following grading for the
tilded oscillators\footnote{ This grading is consistent
with the similar grading of the ``untilded'' oscillators,
if one also chooses $deg(\gamma_0) = deg(\beta_0) = 0$.} with $n \neq 0$:
\beqar
deg(\alpha_n^+) = deg(c_n)=deg({\teta}_n) = deg(\gamma_n) &=& \phantom{-} 1
		\nonumber\\
deg(\alpha_n^-) = deg(\tb_n)=deg(\zeta_n) = deg(\beta_n) &=& -1
		\label{degs}\stop
\eeqar
With this grading, the operator $\hat{d}$ decomposes into
$\hat{d} = \hat{d}_0 + \hat{d}_1 + \hat{d}_2$,
where the $d_i$'s are given by
\beqar
\hat{d}_0 &=& \sum_{n\neq 0} P^+(n)c_{-n}\alpha_n^- +
         \teta_n\beta_{-n}\nonumber\\
\hat{d}_1 &=& \sum_{n,m \neq 0 \atop n+m \neq 0}
c_{-n} \left( \alpha_{-m}^+\alpha_{n+m}^- +
\half (m-n)c_{-m}\tb_{n+m} +    m \zeta_{ -m}\teta_ {n+m} +
m\beta_{-m}\gamma_{n+m} \right)\nonumber\\
\hat{d}_2 &=& \sum_{n\neq 0} P^-(n)c_{-n}\alpha_n^+ \stop \label{dhats}
\eeqar

Following ref.~\cite{Pilch} we distinguish two possible cases of
momenta, and obtain the cohomology of $\hat{d}_0$ from the two
following theorems:
\begin{theorem}
If $P^+(n)\neq 0$ or $P^-(n)\neq 0 ~\forall n \neq 0$, the relative
cohomology of
$\hat{d}_0$ exists only at ghost number 0, and is one-dimensional. These
states in the cohomology are the tachyons, and they satisfy the mass shell
condition $p^+p^-=0$.
\label{th-t}
\end{theorem}
{\bf Proof}: Consider the case $P^+(n)\neq 0~\forall n \neq 0$.
(The proof for the
other case is obtained by reversing the grading of eqs\eq{degs}.)
Define the operator
\beq
K=\sum_{n\neq 0} \left( \f1{P^+(n)}\alpha_{-n}^+\tb_n - n\gamma_n\zeta_{-n}
\right) \stop
\eeq
Since
\beq
\{\hat{d}_0,K\}=\hat{L}_0 \comma
\eeq
any state of level $\hat{L}_0 >0$ which is closed under $\hat{d}_0$ is
also exact. At level zero, since $L_0 = \hat{L}_0 = 0$, the state satisfies
$p^+p^-=0$.  It is clearly in the cohomology of $\hat{d}_0$.  Note that
since all the terms of $\hat{d}_1$ and $\hat{d}_2$ contain oscillator modes,
the tachyons are also in the cohomology of $\hat{d}$.

\begin{theorem}
If  $P^+(r) = P^-(s) = 0$ for non-zero integers $r$ and $s$,
then  $rs>0$ and the relative cohomology of $\hat{d}_0$ is
represented by the following states:
\beqar
\hbox{(i) For $r,s<0$ :}\qquad&
       (\alpha_r^-)^{-s}\ket{p,p_\varphi} & \hbox{and~~}
       \tb_r(\alpha_r^-)^{-s-1}\ket{p,p_\varphi}  \label{l1}\semi\\
\hbox{(ii) For $r,s>0$ :}\qquad&
      (\alpha_{-r}^+)^s\ket{p,p_\varphi} & \hbox{and~~}
       c_{-r}(\alpha_{-r}^+)^{s-1}\ket{p,p_\varphi}  \label{l2} \stop
\eeqar
\label{th-d}
\end{theorem}
{\bf Proof}: If $P^+(r) = P^-(s) = 0$, eq\eq{pn} implies that $p^+p^- = -rs$.
Thus, in order to have states in the kernel of $L_0=\hat{L}_0+p^+ p^-$, one
needs $rs>0$.  Now, define the operator
\beq
K_r=\sum_{n\neq 0,r}\f1{P^+(n)}\alpha_{-n}^+\tb_n - \sum_{n
\neq 0} n \gamma_n \zeta_{-n} \stop \label{Kr}
\eeq
It satisfies:
\beq
\{\hat{d}_0,K_r\}=\hat{L}_{0,r} \comma \label{l0rhat}
\eeq
giving $\hat{L}_{0,r}$---the level operator for all the oscillators except
$b_r$, $c_{-r}$ and $\alpha_r^-$, $\alpha_{-r}^+$.  Since
any state in the cohomology of the theory must now be in the kernel of
$\hat{L}_{0,r}$, it must be generated by these remaining oscillators.
The only such states satisfying the condition $\hat{L}_0 = r s$ are the  ones
written in  (\ref{l1}) and (\ref{l2}).  It is trivial to see that these states
are indeed in the cohomology of $\hat{d}_0$.

\subsection{The full cohomology of $Q_B$}

We now have all the necessary ingredients to classify the
cohomology of $Q_B$.
First, an elementary result of the
cohomology theory of filtered complexes states that {\it
if the cohomology of $\hat{d}_0$ occurs at only one degree, the
cohomology of $\hat{d}$ is in a one to one correspondence with the
cohomology of $\hat{d_0}$\/} \cite{seefor}.
Together with theorems~\ref{th-t} and \ref{th-d}, this immediately implies
that:
\begin{theorem}
The relative cohomology of $\hat{d}$ is given by:\\
(i) The tachyons: A one-dimensional cohomology at ghost number $0$, with
mass-shell condition $p^+ p^- =0$.\\
(ii) A one-dimensional cohomology at ghost numbers $-1$ and $0$ for all
$p^+=r$, $p^-=-s$, with $r$ and $s$ negative integers. \\
(iii) A one-dimensional cohomology at ghost numbers 0 and $+1$ for all
$p^+=r$, $p^-=-s$, with $r$ and $s$ positive integers. \\
\label{th-dhat}
\end{theorem}
We have already seen that the tachyonic states in the cohomology of
$\hat{d}_0$ are in fact states in the full cohomology of $\hat{d}$.
It is less easy to find
explicit representatives for the discrete states in the cohomology, but they
are classified by this theorem.

The full cohomology of $Q_B$ now follows from theorems~\ref{a} and \ref{b}:
Denoting the states of theorem~\ref{th-dhat} by $\ket\psi$, the absolute
cohomology is generated by the states
\goodbreak
\beq
\ket\psi \otimes \ket{{\calp}=0} \otimes\ket{\ua}_\gh
           \otimes\ket{\da}_{bc}
\eeq
\nobreak
and
\nobreak
\beq
a_0 \left( \ket\psi \otimes \ket{{\calp}=0} \otimes\ket{\ua}_\gh
           \otimes\ket{\da}_{bc} \right) \comma
\eeq
\goodbreak
and their conjugates\footnote{Note that in this case, $\ket\psi$ must be
written in terms of the tilded oscillators.}:
\beq
\ket\psi \otimes \ket{\calx = 0}\otimes \ket{\da}_\gh
           \otimes\ket{\da}_{bc}
\eeq
and
\beq
a_0 \left( \ket\psi \otimes \ket{\calx = 0}\otimes \ket{\da}_\gh
           \otimes\ket{\da}_{bc} \right)  \stop
\eeq

\subsection{Comparison to the conformal gauge}

Recall that in studying the spectrum, we can restrict ourselves to
the states built on $\ket{\calp=0}$.
In order to compare the cohomology that we have found to that of
the conformal gauge, it is convenient to change to a more conventional
notation.  First, the tachyon operators in the absolute cohomology can be
written as:
\beq
c \, e ^ { i p x + i p_\varphi^+ \varphi }
\onarrow{a} a \,
c \, e ^ { i p x + i p_\varphi^+ \varphi }
\qquad
c \, e ^ { i p x + i p_\varphi^- \varphi }
\onarrow{a} a \,
c \, e ^ { i p x + i p_\varphi^- \varphi }
\comma
\eeq
with
\beq
p_\varphi^\pm = -i ( \sqrt2 \pm |p| ) \semi
\eeq
they agree with the tachyon states of the conformal gauge.

For the discrete states,
associate the ``$\tb_r$'' states of theorem~\ref{th-d} with the
operators ${\calo}_{u,n}$, the ``$\alpha_r^-$'' states with the operators
$Y_{u+1,n}^+$, the ``$\alpha_{-r}^+$'' states with the operators
$Y_{u+1,n}^-$ and the ``$c_{-r}$'' states with the operators
$P_{u,n}$.  states.  $u$ and $n$ are related to $r$ and $s$ by:
\beqar
u &=& \f{|r+s|}2 -1 \nonumber\\
n &=& \f{r-s}2 \comma
\eeqar
so $u,n \in \ZZ/2$, and $|n| \le u$.  Also note that
the states with $r,s$ negative have $\varphi$ ``momenta'' satisfying the
Seiberg condition \cite{Seiberg}, while those with $r,s$ positive are
anti-Seiberg
states.  The discrete states now fall into the familiar ``diamond''
form of Witten and Zwiebach \cite{Zwie}:
\beq
\matrix{ &  & a{\calo}_{u,n} &  &  \cr
 & \onnearrow{a\ \ \ \ } &  & \searrow &  \cr
{\calo}_{u,n} &  &  &  & aY^+_{u+1,n} \cr
 & \searrow &  & \undernearrow{\ \ a} &  \cr
 &  & Y^+_{u+1,n} &  &  \cr\cr }
\qquad
\matrix{ &  & aY^-_{u+1,n} &  &  \cr
 & \onnearrow{a\ \ \ \ } &  & \searrow &  \cr
Y^-_{u+1,n} &  &  &  & a P_{u,n} \cr
& \searrow &  & \undernearrow{\ \ a} &  \cr
 &  & P_{u,n}  &  &  \cr\cr }
\eeq
and therefore agree with those of the conformal gauge \cite{Lian,Pilch,Witten}.
We thus obtain the main result of this section:

\noindent
{\it The cohomology of the light-cone theory in the
Fock-space representation of the current algebra is equivalent to that of
the Liouville theory.\/}

\heading{The ground ring in the light-cone gauge}

Since the structure of the cohomology of the light-cone gauge (in the Fock
space) is equivalent to that of the conformal gauge, the only problem
remaining to us is to find explicit representatives of the cohomology.
Start with the spin zero, ghost number zero operators $\calo_{u,n}$.
In the conformal gauge, they form a
commutative ``ground ring'' \cite{Witten}, in that under the operator
product expansion:
\beq
O_{s,n}(z)O_{s',n'}(0) \sim O_{s+s',n+n'}(0) \comma
\eeq
up to states exact under $Q_B$.  Denoting $O_{\f12 ,\f12 }$ by $X$ and
$O_{\f12 ,-\f12}$ by $Y$, one sees that the ring is a polynomial ring
generated by $X$ and $Y$, with
\beq
O_{s,n} = X^{s+n}Y^{s-n} \stop
\eeq
Note that $O_{0,0}$ is the identity operator.  In the conformal gauge
$X$ and $Y$ are given by
\beqar
X &=& \left( cb + \f{i}{\sqrt2}(\pa x -i\pa \phi)\right)
         e^{\f{i(x+i\phi)}{\sqrt2}}\nonumber\\
Y &=& \left(cb - \f{i}{\sqrt2}(\pa x +i\pa \phi)\right)
         e^{\f{-i(x-i\phi)}{\sqrt2}} \stop
\eeqar

In the light-cone gauge this structure is reproduced, since the arguments
leading to the existence of the ground ring are unchanged, and the two
cohomologies are in a one-to-one correspondence.  However, before we proceed,
we need
to check that the identity operator $\calo_{0,0}=I$ or, rather, the vacuum
state $\ket0$ is in the cohomology.   The vacuum state occurs at the first
level of the theory, with $r=s=-1$.  In the cohomology of $\hat{d}_0$ at
this level, theorem~\ref{th-d} gives us the state
$\tb_{-1}\ket{p=0,p_\varphi=0}$, which is also in the cohomology of
$\hat{d}$.  Using theorem~\ref{b}, we obtain the relative-cohomology
state\footnote{Note that on all states with $\calp=\gamma_0=0$, one can drop
the ``tildes'' on the oscillators, since the difference between the tilded
and untilded oscillators is proportional to $\gamma_0$ (see eqs\eq{tilding}).}
$b_{-1}\ket{p=0,p_\varphi=0,\gamma_0=0}
\otimes\ket{\ua}_{\gh}\otimes\ket{\da}_{bc}$, which is indeed the vacuum.

The construction of the generators of the ring $X$ and $Y$ is a little more
complicated.  They occur at the second level, with $rs=2$.  The relevant
states of the $\hat{d}_0$ cohomology  are given by
theorem~\ref{th-d}:
\beqar
\hbox{$r=-1,s=-2$ :}\qquad&
\ket{X_0} = \tb_{-1}\alpha_{-1}^-\ket{\f1{\sqrt2},\f{i}{\sqrt2}} \nonumber\\
\hbox{$r=-2,s=-1$ :}\qquad&
\ket{Y_0} = \phantom{\alpha_{-1}}
               \tb_{-2}\ket{\f{-1}{\sqrt2},\f{i}{\sqrt2}} \stop
\eeqar
In order to obtain  the
corresponding states in the cohomology of $\hat{d}$, we use  an
inductive procedure of ref.~\cite{Pilch}, which we briefly illustrate in our
case: Starting with a state ${\psi}_0$ in the cohomology of $\hat{d}_0$,
construct the state ${\psi}_1$ as
\beq
 {\psi}_1  = -\hat{L}_{0,r}^{-1}K_r\hat{d}_1{\psi}_0 \comma
\eeq
where the operators $\hat{L}_{0,r},K_r$ are given in eqs\eq{Kr} and
(\ref{l0rhat}).  $\psi_1$ is
of degree one greater than $\psi_0$.  Continue inductively, defining
\beq
 {\psi}_{k+1}  = -\hat{L}_{0,r}^{-1}K_r(\hat{d}_1{\psi}_k +
\hat{d}_2{\psi}_{k-1}) \stop
\eeq
Then $\psi = {\psi}_0 + {\psi}_1 + {\psi}_2 + \cdots$ is in the cohomology of
$\hat{d}$.  In our case the procedure stops after two steps, giving the states
\beqar
\ket{X} = \left( -\tb_{-2} + \tb_{-1} \alpha_{-1}^- - \zeta_{-1} \gamma_{-1}
\right)
           \ket{\f1{\sqrt2},\f{i}{\sqrt2}} \nonumber\\
\ket{Y} = \left(\tb_{-2} + \alpha_{-1}^+\tb_{-1} +\zeta_{-1} \gamma_{-1}
\right)
           \ket{\f{-1}{\sqrt2},\f{i}{\sqrt2}} \stop
\eeqar
Using theorem~\ref{b} to turn these states into relative-cohomology states,
we see that the ground ring in the light-cone gauge is generated by the
operators:
\beqar
X^{lc} &=& \left( cb + \pa\alpha^- + c (\pa \zeta ) \gamma \right) \,
          e^{\f{i(x+i\varphi)}{\sqrt2}}
\nonumber\\
Y^{lc} &=& \left( cb - \pa\alpha^+ + c (\pa \zeta ) \gamma \right)   \,
          e^{\f{-i(x-i\varphi)}{\sqrt2}} \stop
\eeqar
These operators have extra pieces depending on $\zeta$ and $\gamma$ compared to
$X$ and $Y$ of the
conformal gauge.  Since the $(\beta,\gamma)$ and $(\zeta,\eta)$ fields
appear here, the light-cone theory can not be regarded
simply as the Liouville theory plus a topological theory.

\heading{Higher ghost number states, and the chiral $w_\infty$ algebra}

We can continue in the same manner to study the higher ghost number
states.  Since, by theorem~\ref{a}, the operator $a$ needed to obtain
the absolute cohomology has the same form
in the light-cone as in the conformal gauge, we can restrict ourselves
to studying the relative cohomology operators $Y_{j,m}^\pm$ and
$P_{u,n}$.  The usual way of constructing the $Y_{j,m}^\pm$ operators
in the conformal gauge is to start with the extra primary states of
the  $c=1$ conformal field theory of a scalar compactified on a circle
with self-dual radius $R = \sqrt2$, $V_{j,m}$ \cite{Kac}, and to dress
them with the Liouville field to produce the operators $W_{j,m}^\pm$:
\beq
W_{j,m}^{\pm}=V_{j,m} e^{(\sqrt2(1\mp j)\phi) } \stop
\eeq
As was shown
in \cite{Kleb}, the charges
$Q_{j,m}^+=\f1{2{\pi}i}\oint W_{j,m}^+$ satisfy a chiral $w_{\infty}$ algebra.
One can then form the desired BRST invariant operators $Y_{j,m}^\pm$:
\beq
Y_{j,m}^{\pm}=cW_{j,m}^{\pm} \stop
\eeq
However, one can also calculate these operators directly from the
$\hat{d}_0$ cohomology states, using the method illustrated in the previous
section.  Note that the states of the light-cone $\hat{d}_0$ cohomology
have the same form as those of the conformal gauge in ref.~\cite{Pilch}.
Also, except for
the ground-ring states, the new light-cone oscillators
$\beta_n,\gamma_n,\zeta_n,\eta_n$ do not appear when $\hat{d}_1$ and
$\hat{d}_2$ act
on the states.  The procedure therefore gives exactly the same result
in the light-cone as in the conformal gauge, and the operators
$Y_{j,m}^\pm$ that we had previously are representatives of the
cohomology in the light-cone gauge.  The ghost number two operators $P_{u,n}$
are also the same as those of the conformal gauge.

\section{The Kac-Moody BRST cohomology from Felder's resolution}

\subsection{The Kac-Moody and Fock-space cohomologies at the first level}
\label{51}

We can now return to the problem of obtaining the full cohomology on the
irreducible Kac-Moody modules, from the cohomology on the Fock space.
First, in order to illustrate the issues involved in this procedure, we
shall examine both of these cohomologies
at the first level, restricting ourselves to the relative cohomology of
$\hat{Q}$.  The states in the Fock space are easily derived using the results
of the previous section, and are summarized in table~1\footnote{From
this section on, we measure ghost numbers from the vacuum state $\ket0$.}.
As was shown in the
previous section, the light-cone Fock-space states are equivalent to those of
the conformal gauge; in fact, at this level they are identical.   The states of
the relative cohomology of the current algebra at the first level are derived
in appendix~A, and the states built on lowest weight states (LWS) are given
in table~2. (All the states in the cohomology at level 1 have $p=0$,
so we shall suppress $p$ in this section, and write only $p_\varphi$ or
$J$ and $M$.  We also suppress the $\ket\da_{bc}$.)   Finally, we show the
conjugate states of the Fock-space theory in table~3, and the HWS of the
current algebra in table~4.

\begin{table}
\begin{center}
\renewcommand{\arraystretch}{1.5}
\begin{tabular}{|c||c|c|}
\hline
\strut
ghost number &
state $\left( \ket{p_\varphi} \otimes\ket{\gamma_0=0} \upgh \right) $ &
operator \\
\hline
0  & $b_{-1}\ket{0}$ & $I$\\
\hline
1  &  $x_{-1}\ket{0} $ & $c \pa x$ \\
\cline{2-3}
   &  $x_{-1}\ket{-2i\sqrt2} $ & $c \pa x e^{2\sqrt2\varphi}$ \\
\hline
2  &  $c_{-1}\ket{-2 i\sqrt2}$ &
            $c \pa c e^{2\sqrt2\varphi}$ \\
\hline
\end{tabular}
\renewcommand{\arraystretch}{1.0}
\caption{First-level states in the Fock space.}
\end{center}
\label{tab}
\end{table}
\begin{table}
\begin{center}
\renewcommand{\arraystretch}{1.5}
\begin{tabular}{|c||c|}
\hline
\strut
ghost number &
state $\left( \ket{J,M} \upgh \right) $ \\
\hline
0  & $b_{-1}\ket{-\thrhalf ,\thrhalf } $ \\
\hline
1  &  $x_{-1}\ket{-\thrhalf , \thrhalf } $ \\
\cline{2-2}
   &  $ x_{-1}\ket{\half ,-\half }$ \\
\hline
2  &  $c_{-1}\ket{\half ,-\half }$ \\
\cline{2-2}
   &  $\eta_{-1}\ket{-\thrhalf ,\thrhalf }$ \\
\hline
\end{tabular}
\renewcommand{\arraystretch}{1.0}
\caption{First-level states in the current algebra.}
\end{center}
\end{table}

\begin{table}
\begin{center}
\renewcommand{\arraystretch}{1.5}
\begin{tabular}{|c||c|}
\hline
\strut
ghost number &
state $\left( \ket{p_\varphi} \otimes\ket{\beta_0=0} \downgh \right) $ \\
\hline
-1  & $(b_{-1}-\zeta_{-1} \gamma_0) \ket{0}$ \\
\hline
0  &  $x_{-1}\ket{0}$ \\
\cline{2-2}
   &  $x_{-1}\ket{-2i\sqrt2}$ \\
\hline
1  &  $c_{-1}\ket{-2 i\sqrt2}$ \\
\hline
\end{tabular}
\renewcommand{\arraystretch}{1.0}
\caption{First-level conjugate states in the Fock space.}
\end{center}
\end{table}
\begin{table}
\begin{center}
\renewcommand{\arraystretch}{1.5}
\begin{tabular}{|c||c|}
\hline
\strut
ghost number &
state $\left( \ket{J,M} \downgh \right) $ \\
\hline
-1  & $\zeta_{-1}\ket{-\thrhalf ,-\thrhalf }$ \\
\cline{2-2}
    & $( b_{-1}- \zeta_{-1} J_0^+ ) \ket{\half ,\half } $ \\
\hline
 0  &  $x_{-1}\ket{\half ,\half }$ \\
\cline{2-2}
    &  $x_{-1}\ket{-\thrhalf , -\thrhalf }$ \\
\hline
 1  &  $c_{-1}\ket{-\thrhalf ,-\thrhalf }$ \\
\hline
\end{tabular}
\renewcommand{\arraystretch}{1.0}
\caption{First-level conjugate states in the current algebra.}
\end{center}
\end{table}

In order to compare the current algebra to the Fock space,
one needs the relation between the $\ket{J,M}$ states and the
``Liouville'' states $\ket{p_\varphi}$.  For generic\footnote{If $2J+1
\in \NN$, the HWS and LWS are both in the same finite $SU(2)$
representation so this identification is not unique.  We shall see the
importance of this later.} HWS and LWS, this is
determined from the definitions of the currents in eqs\eq{wakimoto} to be:
\beqar
\ket{J,J} &\leftrightarrow& \ket{p_\varphi =
\sqrt2 i (J-\half) } \otimes\ket{\beta_0=0} \nonumber\\
\ket{J,-J} &\leftrightarrow& \ket{p_\varphi =
\sqrt2 i (-J-\thrhalf) } \otimes\ket{\gamma_0=0} \stop \label{phiJ}
\eeqar
Using these relations,
one sees from the tables that the states in the current algebra
are in a one-to-one correspondence with those of the Fock space (or
equivalently with those of the conformal gauge), except for the extra
pair of conjugate states $\eta_{-1}\ket{-\thrhalf ,\thrhalf } \upgh$
and $\zeta_{-1}\ket{-\thrhalf ,-\thrhalf }\downgh$ in the current
algebra.  The existence of these extra states at the first level
indicates that
the BRST cohomology on
the irreducible $SL(2,R)$ Kac-Moody modules is inequivalent to
that of the conformal gauge.  However, since there are an infinite number of
discrete states in both cohomologies, we can not yet draw a definite
conclusion.  This will be possible after we obtain the full Kac-Moody
cohomology in section~\ref{full}.

\heading{From free fields to current algebra at the first level}

In order to understand why extra states appeared in the current
algebra formulation, it is useful to recall some facts about
free-field representations of theories:
A free field description of a conformal field theory provides a
realization of the chiral algebra---in our case the $SL(2,R)$ Kac-Moody
algebra---by free fields.  In addition, one needs to have
a projection from the free-field Fock space to the irreducible
representations of the chiral algebra.  This projection is essentially
a null states decoupling, and needs to take care of two problems that
are dual to each other :\\
(i) Singular states: Null states in the Kac-Moody module do not
necessarily vanish in the free-field Fock space.\\
(ii) Cosingular states: There can exist states in the Fock space that do not
have analogous states in the irreducible Kac-Moody module.

The extra states that we have found in the first level relative
cohomology illustrate both cases: The state
$\zeta_{-1}\ket{-\thrhalf ,-\thrhalf }\otimes\ket{ \da}_{\gh}$
is closed, since
\beq
\hat{Q} \left ( \zeta_{-1}\ket{-\thrhalf ,-\thrhalf } \downgh \right ) =
J_{-1}^-\ket {-\thrhalf ,-\thrhalf }\downgh  \comma
\eeq
and $J_{-1}^-\ket{-\thrhalf ,-\thrhalf}$ is a null state, being
annihilated by all $J_{n >0}^a$'s.  However, acting $\hat{Q}$ on the equivalent
state in the Fock-space realization, one obtains
\beq
\hat{Q} \left ( \zeta_{-1}\ket{ p_\varphi = -2 \sqrt2 i , \beta_0=0}
\downgh \right ) =
\beta_{-1} \ket{ p_\varphi = -2 \sqrt2 i , \beta_0=0} \downgh \comma
\label{frsing}
\eeq
which is a nonvanishing singular vector.
Thus the state is not closed in the Fock-space
analysis, and one has an extra state in the current algebra.

In the conjugate case, the analogue of the extra state
$\eta_{-1}\ket{-\thrhalf ,\thrhalf }$ in the Fock space is exact:
\beq
\eta_{-1}\ket{p_\varphi=0 , \gamma_0=0} \upgh =
\hat{Q} \left( \gamma_{-1}\ket{p_\varphi=0 , \gamma_0=0} \upgh \right)
\label{frcos}
\stop
\eeq
However, one cannot construct the analogue of the cosingular state
$\gamma_{-1}\ket{p_\varphi=0, \gamma_0=0}$ in the Kac-Moody module, since
\beq
\gamma_{-1} \ket{p_\varphi ,  \gamma_0=0}
= \f{i}{\sqrt2 p_\varphi} J_{-1}^+ \ket{p_\varphi , \gamma_0=0}
\eeq
becomes ill defined as $p_\varphi \to 0$.
Thus, again, an extra state appears in the analysis on the Kac-Moody module.

\subsection{The Felder resolution}

We have seen at the first level that the BRST cohomology on the
irreducible $SL(2,R)$ Kac-Moody module is different than the BRST
cohomology on the free field Fock space.  The general mechanism for
obtaining cohomologies on irreducible modules of chiral
algebras from those of free field Fock spaces is called a Felder resolution
\cite{Felder}.  It is needed whenever the Fock-space module $F_\Lambda$
carries a reducible representation of the chiral algebra.
One then needs a projection from the free-field Fock space to the
irreducible representations of the chiral algebra.  This is done by a
BRST like procedure: First one needs operators $Q^{(i)}$ that commute with
the chiral algebra.  These operators are built out of screening operators.
They change
the momentum of the Fock space, giving a map between different modules
$F_\Lambda^{(i)}$, with $(i)$ denoting the Felder index.  One then has
a complex,
\marginnote{I removed your complex!}
which is
generically of the form:
\beq
\cdots \To F_\Lambda^{(-1)} \OnArrow{Q^{(-1)}} F_\Lambda^{(0)}
\OnArrow{Q^{(0)}} F_\Lambda^{(1)} \To \cdots \stop \label{complex}
\eeq
Since the $Q^{(i)}$'s commute with the chiral algebra, the cohomology
groups of the complex are chiral algebra modules.  The cohomology is
nontrivial only at one Felder index, chosen to be 0, where it is
isomorphic to the irreducible representation $L_\Lambda$ of the chiral
algebra.

\setlength{\unitlength}{.6cm}
\thicklines

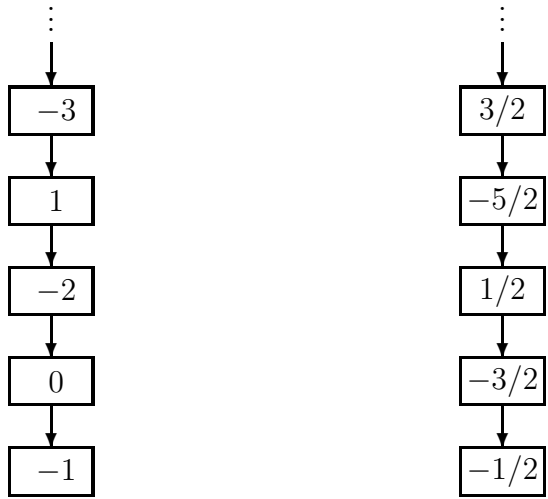
\begin{figure}
\begin{center}
\begin{picture}(15,11.5)(-7.5,-1.5)

	\put(-5,9.2){\makebox(0,0){                           $\vdots$ }}
		\put(-5,8.5){\vector(0,-1){1}}
	\put(-5,7){\makebox(0,0){\framebox(1.8,1){           $-3$}}}
		\put(-5,6.5){\vector(0,-1){1}}
	\put(-5,5){\makebox(0,0){\framebox(1.8,1){           $1$}}}
		\put(-5,4.5){\vector(0,-1){1}}
	\put(-5,3){\makebox(0,0){\framebox(1.8,1){           $-2$}}}
		\put(-5,2.5){\vector(0,-1){1}}
	\put(-5,1){\makebox(0,0){\framebox(1.8,1){           $0$}}}
		\put(-5, .5){\vector(0,-1){1}}
	\put(-5,-1){\makebox(0,0){\framebox(1.8,1){           $-1$}}}

	\put(5,9.2){\makebox(0,0){                           $\vdots$ }}
		\put(5,8.5){\vector(0,-1){1}}
	\put(5,7){\makebox(0,0){\framebox(1.8,1){            $3/2$ }}}
		\put(5,6.5){\vector(0,-1){1}}
	\put(5,5){\makebox(0,0){\framebox(1.8,1){            $-5/2$ }}}
		\put(5,4.5){\vector(0,-1){1}}
	\put(5,3){\makebox(0,0){\framebox(1.8,1){            $1/2$ }}}
		\put(5,2.5){\vector(0,-1){1}}
	\put(5,1){\makebox(0,0){\framebox(1.8,1){            $-3/2$ }}}
		\put(5, .5){\vector(0,-1){1}}
	\put(5,-1){\makebox(0,0){\framebox(1.8,1){            $-1/2$ }}}

\end{picture}
\end{center}
\caption{The basic inclusion diagrams of the $\kappa=-3$ $A_1^{(1)}$ Kac-Moody
modules}
\end{figure}

In our case the chiral algebra is an $A_1^{(1)} \sim SL(2,R)$ Kac-Moody
algebra of level \hbox{$\kappa=-3$}.  Representations are characterized
by their vacuum, either a HWS or a LWS with spin $J$.   When
$\kappa= -3$, one can see from the Kac-Kazhdan formula \cite{KK}
that representations can be reducible only when $J \in \ZZ/2$.  The
basic inclusion diagrams, illustrating which representations are
contained in each other, are given in figure~1.  We shall denote the Fock
spaces modules built on $\ket{\gamma_=0}$ by $F_J$, and their conjugate
modules built on $\ket{\beta_=0}$ by $F_J^*$, where $J$ is determined from
$p_\varphi$ by eqs\eq{phiJ}.  Recall that the vacuum state of $F_J$ is a
LWS, while that of $F_J^*$ is a HWS.  The BRST operators are built from the
basic screening operator which, in our case, is given by:
\beq
V \equiv \beta e^{\sqrt2 \varphi} \stop
\eeq
The operator $Q_n$ is made by taking appropriate contour integrals of
the product of $n$ $V$'s \cite{Felder}.  Thus, as can be seen from
eq\eq{phiJ}, it raises the $J$'s of the Fock spaces, $Q_n : F_J \to
F_{J+n}$, and lowers the $J$'s of the conjugate Fock spaces, $Q_n :
F_J^* \to F_{J-n}^*$.

In order to obtain the resolution of the algebra, we now
need to build the Felder complex. The Felder
resolution for $A_1^{(1)}$ has
been carried out for the cases where $\kappa +2$ is a ``generic''
complex number, giving an inclusion diagram with just two representations,
and when $\kappa+2$ is a positive rational number, giving a double-branch
inclusion diagram \cite{BFelder}.  It has
not been carried out for our degenerate single-branch case.  However, the
Felder resolutions of the Virasoro algebra and of the
$A_1^{(1)}$ Kac-Moody algebra
share many properties, and the resolution
of the single-branch Virasoro case has been carried out\footnote{We would
like to thank Profs.  P.~Bernard and G.~Felder for e-mail concerning this
issue.} \cite{1branch}.  In that case it was found that the complex reduces to
short complexes of the type $0\rightarrow A \rightarrow B\rightarrow 0$.
We shall therefore assume that such a structure appears in our case.
Examining the
form of the inclusion diagrams of figure~1, this leads us to the following
conjecture for the Felder complex:

\begin{qtheorem}
\label{qth}
For $J \in \ZZ/2$, $J \ne -\half, -1$, the Felder complex
of the $\kappa=-3$ $A_1^{(1)}$ Kac-Moody algebra is:
\beq
\eqalign{
\hbox{$J \ge 0$ :}\qquad&
      0 \To F_{-J-1}^{(-1)} \OnArrow{Q_{2J+1}} F_J^{(0)} \To 0  \cr
\hbox{$J \le -\thrhalf$ :}\qquad&
      0 \To F_J^{(0)} \OnArrow{Q_{-2J-2}} F_{-J-2}^{(1)} \To 0  \comma
}
\eeq
with the dual complex:
\beq
\eqalign{
\hbox{$J \ge 0$ :}\qquad&
      0 \To F_J^{*(0)} \OnArrow{Q_{2J+1}} F_{-J-1}^{*(-1)} \To 0 \cr
\hbox{$J \le -\thrhalf$ :}\qquad&
      0 \To F_{-J-2}^{*(-1)} \OnArrow{Q_{-2J-2}} F_J^{*(0)} \To 0  \stop
}
\eeq
For all other $J$'s the complex is trivial.
The only nontrivial cohomology of the complex occurs at index
$i=0$, and is isomorphic to the irreducible Kac-Moody module $\Lambda_J$.
\end{qtheorem}

We have not proven this conjecture, but have checked that it gives
the correct results on Fock spaces with small $J$'s.  In particular
the conjectured Felder cohomology takes care of the leading singular vector
$(\beta_0)^{2J+1} \ket{J,-J}$ for $J \ge 0$, and the leading cosingular
vector $(\gamma_{-1})^{-2J-2} \ket{J,-J}$ for $J \le -\thrhalf$.

\subsection{The Kac-Moody cohomology}
\label{full}

The relative cohomology of $\hat{Q}$ acting on the current algebra at
ghost number $n$ is given by $H_{rel}^n (\Lambda_J) = H_{rel}^n (H_F^0)$.
This cohomology can be evaluated using the theorem:
\begin{theorem}
If the Felder complex exists only at degrees 0 and either $+1$ or $-1$, then
$H_{rel}^n (H_F^0) = H_F^0 (H_{rel}^n) + H_F^{\pm 1} ( H_{rel}^{n \mp 1} )$.
\label{spect}
\end{theorem}
The proof is given in appendix~B.

In order to see the implications of this theorem, we need to find  $H_F^0
(H_{rel}^n)$ and $ H_F^{\pm 1} ( H_{rel}^{n \mp 1} )$.  $H_{rel}^n$ consists
of the states of ghost number $n$ that
we have found previously in the cohomology of the Fock space.
Denote such states  by $\Psi_J^n$.  One can
simplify the analysis by noting that, except for the ``ground-ring''
states, these $\Psi$'s are all in the vacuum state of the Felder Fock space,
\ie{} they contain no oscillator modes of the Wakimoto fields.
Also, the ground-ring states do not have to be considered separately, since
their conjugates are Felder vacuum states.

Now, consider first a Felder
complex of the form  $0\rightarrow F_J^{(0)} \onarrow{Q_F} F_{J'}^{(1)}
\rightarrow 0$.  In this case,
$H_F^0 (H_{rel}^n)$ consists of the $\Psi_J^n$'s that are
annihilated by $Q_F$.  Since the $\Psi$'s are vacuum vectors, and are
therefore not cosingular, the cohomology consists simply of the states in
the current algebra corresponding  directly to the $\Psi_J^n$'s.  In order
to find states of  $H_F^1 ( H_{rel}^{n- 1} )$, one needs to pull back (the
vacuum state)  $\Psi_{J'}^{n-1}$ to a cosingular vector $\chi_J^{n-1}$,
satisfying $Q_F \, \chi_J^{n-1} = \Psi_{J'}^{n-1}$. This can always
be done since $Q_F$ maps
$F_J$ onto $F_{J'}$.  The desired state in the current-algebra cohomology is
then the state corresponding to $\hat{Q}$ acting on the cosingular
vector $\chi_J^{n-1}$. Schematically,
\beq
\Psi_{J}^n \sim \hat{Q}\, Q_F^{-1} \Psi_{J'}^{n-1} \stop
\eeq
Note that this
procedure has increased the ghost number of the state, as required by
theorem~\ref{spect}.
As an example of such a state, start with the LWS cosmological constant state
\beq
\Psi_{-1/2}^1 = \ket{\gamma_0=0,p_\varphi=0,p=0}\upgh \stop
\eeq
One can then find the cosingular state
\beq
\chi_{-3/2}^1 = \gamma{-1} \ket{\gamma_0=0,p_\varphi=0,p=0}\upgh
\eeq
satisfying $Q_1 \chi_{-3/2}^1 = \Psi_{-1/2}^1$.  As seen in eq\eq{frcos},
\beq
\Psi_{-3/2}^2 = \hat{Q} \, \chi_{-3/2}^1
\eeq
then corresponds to the extra
first-level state $\eta_{-1}\ket{J=-\thrhalf ,M=\thrhalf,p=0} \upgh$ found in
section~\ref{51}.


\setlength{\unitlength}{2cm}
\thicklines

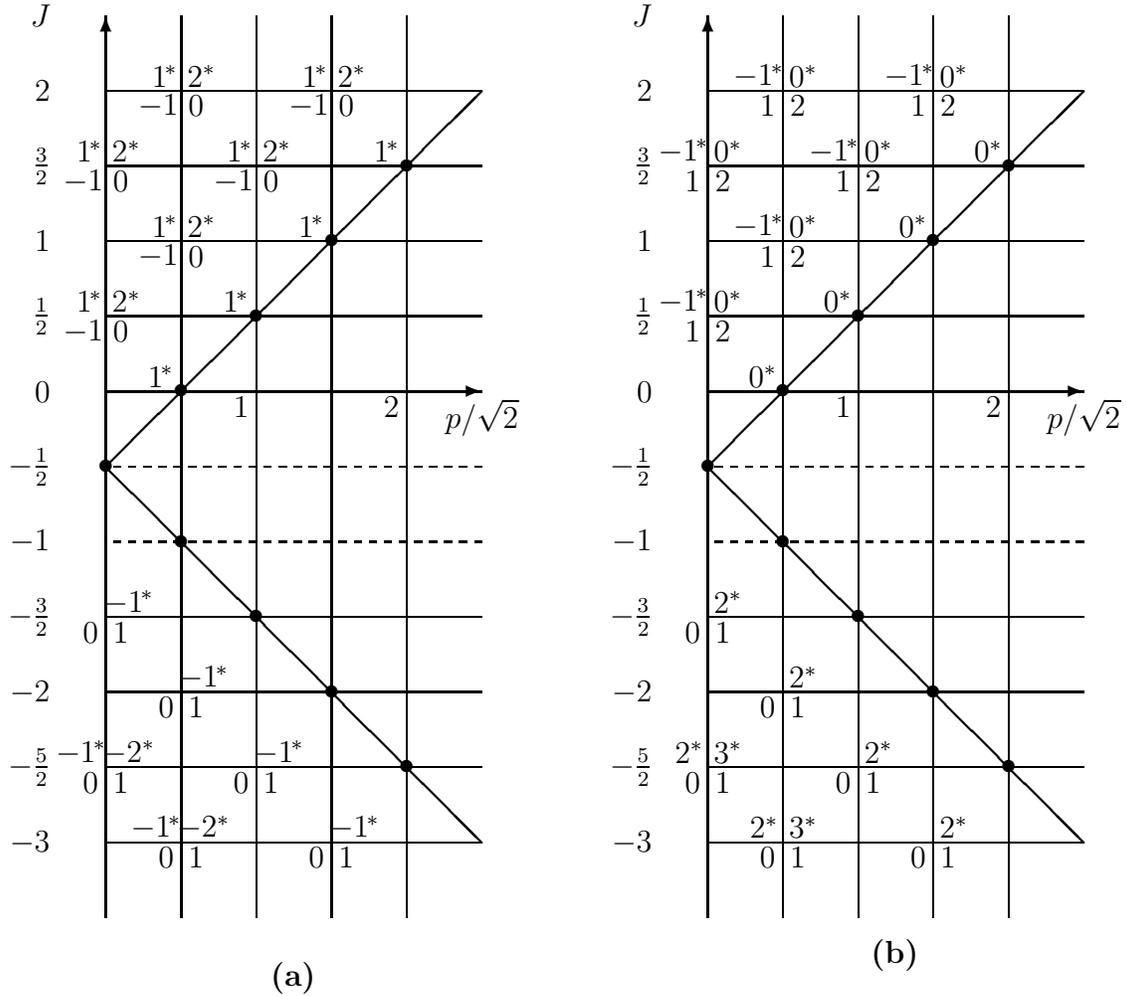
\begin{figure}
\begin{center}
\begin{picture}(8,7)(0,0)

\put(0,0){
\begin{picture}(4,7)(-1,-4)

\put(0,-3.5){\vector(0,1){6}}
\put(0,0){\vector(1,0){2.5}}
\put(0,-.5){\line(1,1){2.5}}
\put(0,-.5){\line(1,-1){2.5}}

\put(-.6,2.5){\makebox(.3,0)[r]{ $J$ }}
\put(2.5,0){\makebox(0,-.05)[t]{$p/\sqrt2$}}
\put(1.25,-3.9){\makebox(0,0){ \bf (a) }}

\thinlines
\multiput(0,-3)(0,.5){4}{\line(1,0){2.5}}
\multiput(0,.5)(0,.5){4}{\line(1,0){2.5}}
\multiput(0.5,-3.5)(.5,0){4}{\line(0,1){6}}
\multiput(.05,-1)(.1,0){25}{\line(1,0){.05}}
\multiput(.05,-.5)(.1,0){25}{\line(1,0){.05}}

\put(-.6,-3){\makebox(.3,0)[r]{ $-3$ }}
\put(-.6,-2.5){\makebox(.3,0)[r]{ $-\f52$ }}
\put(-.6,-2){\makebox(.3,0)[r]{ $-2$ }}
\put(-.6,-1.5){\makebox(.3,0)[r]{ $-\f32$ }}
\put(-.6,-1){\makebox(.3,0)[r]{ $-1$ }}
\put(-.6,-.5){\makebox(.3,0)[r]{ $-\f12$ }}
\put(-.6,0){\makebox(.3,0)[r]{ $0$ }}
\put(-.6,.5){\makebox(.3,0)[r]{ $\f12$ }}
\put(-.6, 1){\makebox(.3,0)[r]{ $1$ }}
\put(-.6,1.5){\makebox(.3,0)[r]{ $\f32$ }}
\put(-.6, 2){\makebox(.3,0)[r]{ $2$ }}

\put(1,0){\makebox(-.2,-.2){ $1$ }}
\put(2,0){\makebox(-.2,-.2){ $2$ }}

\multiput(0,-.5)(.5,.5){5}{\makebox(0,0){ $\bullet$ }}
\multiput(.5,0)(.5,.5){4}{\makebox(-.27,.2){ $1^{\n2 *}$ }}
\multiput(0,.5)(.5,.5){4}{\makebox(-.3,-.2){ $-1$ }}
\multiput(0,.5)(.5,.5){4}{\makebox(.2,-.2){ $0$ }}
\multiput(0,.5)(.5,.5){4}{\makebox(-.22,.2){ $1^{\n2 *}$ }}
\multiput(0,.5)(.5,.5){4}{\makebox(.27,.2){ $2^*$ }}
\multiput(0,1.5)(.5,.5){2}{\makebox(-.3,-.2){ $-1$ }}
\multiput(0,1.5)(.5,.5){2}{\makebox(.2,-.2){ $0$ }}
\multiput(0,1.5)(.5,.5){2}{\makebox(-.22,.2){ $1^{\n2 *}$ }}
\multiput(0,1.5)(.5,.5){2}{\makebox(.27,.2){ $2^*$ }}

\multiput(.5,-1)(.5,-.5){4}{\makebox(0,0){ $\bullet$ }}
\multiput(0,-1.5)(.5,-.5){4}{\makebox(-.2,-.2){ $0$ }}
\multiput(0,-2.5)(.5,-.5){2}{\makebox(-.2,-.2){ $0$ }}
\multiput(0,-1.5)(.5,-.5){4}{\makebox(.2,-.2){ $1$ }}
\multiput(0,-2.5)(.5,-.5){2}{\makebox(.2,-.2){ $1$ }}
\multiput(0,-1.5)(.5,-.5){4}{\makebox(.3,.2){ $-1^{\n2 *}$ }}
\multiput(0,-2.5)(.5,-.5){2}{\makebox(.3,.2){ $-2^*$ }}
\multiput(0,-2.5)(.5,-.5){2}{\makebox(-.4,.2){ $-1^{\n2 *}$}}

\end{picture}
}

\put(4,0){
\begin{picture}(4,7)(-1,-4)

\put(0,-3.5){\vector(0,1){6}}
\put(0,0){\vector(1,0){2.5}}
\put(0,-.5){\line(1,1){2.5}}
\put(0,-.5){\line(1,-1){2.5}}

\put(-.6,2.5){\makebox(.3,0)[r]{ $J$ }}
\put(2.5,0){\makebox(0,-.05)[t]{$p/\sqrt2$}}
\put(1.25,-3.75){\makebox(0,0){ \bf (b) }}

\thinlines
\multiput(0,-3)(0,.5){4}{\line(1,0){2.5}}
\multiput(0,.5)(0,.5){4}{\line(1,0){2.5}}
\multiput(0.5,-3.5)(.5,0){4}{\line(0,1){6}}
\multiput(.05,-1)(.1,0){25}{\line(1,0){.05}}
\multiput(.05,-.5)(.1,0){25}{\line(1,0){.05}}

\put(-.6,-3){\makebox(.3,0)[r]{ $-3$ }}
\put(-.6,-2.5){\makebox(.3,0)[r]{ $-\fhalf$ }}
\put(-.6,-2){\makebox(.3,0)[r]{ $-2$ }}
\put(-.6,-1.5){\makebox(.3,0)[r]{ $-\thrhalf$ }}
\put(-.6,-1){\makebox(.3,0)[r]{ $-1$ }}
\put(-.6,-.5){\makebox(.3,0)[r]{ $-\half$ }}
\put(-.6,0){\makebox(.3,0)[r]{ $0$ }}
\put(-.6,.5){\makebox(.3,0)[r]{ $\half$ }}
\put(-.6, 1){\makebox(.3,0)[r]{ $1$ }}
\put(-.6,1.5){\makebox(.3,0)[r]{ $\thrhalf$ }}
\put(-.6, 2){\makebox(.3,0)[r]{ $2$ }}

\put(1,0){\makebox(-.2,-.2){ $1$ }}
\put(2,0){\makebox(-.2,-.2){ $2$ }}

\multiput(0,-.5)(.5,.5){5}{\makebox(0,0){ $\bullet$ }}
\multiput(.5,0)(.5,.5){4}{\makebox(-.27,.2){ $0^*$ }}
\multiput(0,.5)(.5,.5){4}{\makebox(-.2,-.2){ $1$ }}
\multiput(0,.5)(.5,.5){4}{\makebox(.2,-.2){ $2$ }}
\multiput(0,.5)(.5,.5){4}{\makebox(-.32,.2){ $-1^{\n2 *}$ }}
\multiput(0,.5)(.5,.5){4}{\makebox(.27,.2){ $0^*$ }}
\multiput(0,1.5)(.5,.5){2}{\makebox(-.2,-.2){ $1$ }}
\multiput(0,1.5)(.5,.5){2}{\makebox(.2,-.2){ $2$ }}
\multiput(0,1.5)(.5,.5){2}{\makebox(-.32,.2){ $-1^{\n2 *}$ }}
\multiput(0,1.5)(.5,.5){2}{\makebox(.27,.2){ $0^*$ }}

\multiput(.5,-1)(.5,-.5){4}{\makebox(0,0){ $\bullet$ }}
\multiput(0,-1.5)(.5,-.5){4}{\makebox(-.2,-.2){ $0$ }}
\multiput(0,-2.5)(.5,-.5){2}{\makebox(-.2,-.2){ $0$ }}
\multiput(0,-1.5)(.5,-.5){4}{\makebox(.2,-.2){ $1$ }}
\multiput(0,-2.5)(.5,-.5){2}{\makebox(.2,-.2){ $1$ }}
\multiput(0,-1.5)(.5,-.5){4}{\makebox(.27,.2){ $2^*$ }}
\multiput(0,-2.5)(.5,-.5){2}{\makebox(.27,.2){ $3^*$ }}
\multiput(0,-2.5)(.5,-.5){2}{\makebox(-.25,.2){ $2^*$ }}

\end{picture}
}

\end{picture}
\end{center}

\caption{The relative-cohomology states in the current algebra built on (a) the
HWS and (b) the LWS of the Fock space.  Only the $p\ge0$ sector is
shown.  Discrete states are labeled by the ghost number of
the operators, discrete tachyons by ``blobs''.  The starred states are the
states obtained nontrivially by the Felder resolution.
Reflections through the dashed lines at $J=-\half$
and $J=-1$ give the $J$'s related by the Felder operation for positive and
negative $J$'s, respectively.}

\label{plot}
\end{figure}

In the cases when the Felder complex is of the form  $0\rightarrow
F_{J'}^{(-1)} \onarrow{Q_F} F_J^{(0)} \rightarrow 0$, $H_F^0 (H_{rel}^n)$
again consists  of the states in the current algebra corresponding  to the
$\Psi_J^n$'s, since none of the $\Psi$'s are singular  vectors.  For $H_F^1
( H_{rel}^{n- 1} )$, one maps the state $\Psi_{J'}^n$ to the singular vector
$\chi_J^n = Q_F \Psi_{J'}^n$.  Since this singular vector is not in the
Fock-space cohomology, it can be written as $\chi_J^n = \hat{Q}
\Psi_{J}^{n-1}$, and $\Psi_{J}^{n-1}$ corresponds to the desired state.  Thus
\beq
\Psi_{J}^{n-1} \sim \hat{Q}^{-1} Q_F \Psi_{J'}^n \comma
\eeq
and the ghost number of
the state has been reduced by 1, as required by theorem~\ref{spect}.
As an example of this type, start with the HWS cosmological constant state
\beq
\Psi_{-1/2}^0 = \ket{ \beta_0=0 , p_\varphi = - \sqrt2 i , p=0} \downgh \stop
\eeq
Acting
on it by $Q_1$, one finds the singular vector
\beq
\chi_{-3/2}^0 = \beta_{-1} \ket{ \beta_0=0 , p_\varphi = -2 \sqrt2 i , p=0}
                     \downgh \stop
\eeq
As seen in eq\eq{frsing}, solving $\hat{Q} \Psi_{-3/2}^{-1} = \chi_{-3/2}^0 $
gives one the analogue of the extra first-level state
$\zeta_{-1}\ket{J=-\thrhalf ,M=-\thrhalf , p=0}\otimes\ket{ \da}_{\gh}$
of section~\ref{51}.

\heading{The complete relative cohomology on the current algebra}

In order to keep track of this procedure for arbitrary $J$,
it is convenient to plot the
starting Fock-space cohomology with respect to $p$ and $J(p_\varphi)$.
This is done in figure~\ref{plot}, to which the reader is
referred.   There, the unstared states and the tachyons
refer to the states of the relative Fock-space cohomology.
Graph~2a contains the HWS states in the sector $\ket{\beta_=0}$;
graph~2b the LWS states in the sector $\ket{\gamma_0=0}$.
We now use the
Felder procedure to obtain the states of the current algebra.
As was argued above, all the Fock-space states have direct analogues
in the current algebra, but there are other states in addition.

Consider first the case $J \ge 0$.  The alert reader may have noticed
that we found no new states of this form in the cohomology of the
current algebra at the first level.  That this persists to all
orders can be seen most easily by considering the HWS states in
graph~2a.  We remind the reader that these states are of the form
$\ket{\downarrow}_\gh$.  In this case,
the $J\ge0$ Felder complex of theorem~\ref{qth} is of the form
$0\rightarrow F_J^{*(0)} \onarrow{Q_F} F_{-J-1}^{*(1)} \rightarrow 0$.
Thus one can find new states $\Psi_J^n$ from the states $\Psi_{-J-1}^{n-1}$,
which are found by reflecting $\Psi_J^n$ through the line $J=-\half$ in the
graph. The cosingular states $\chi_J^{n-1}$ satisfying
\beq
Q_{2J+1} \, \chi_J^{n-1}=\Psi_{-J-1}^{n-1}
\eeq
can be taken to be simply
\beq
\chi_J^{n-1}= \gamma_0^{2J+1} e^{-\sqrt2 (2J+1) \varphi} \, \Psi_{-J-1}^{n-1}
\stop
\eeq
Since $[ \hat{Q},\gamma_0 ] = -\eta_0$,
the desired state is given by:
\beqar
\Psi_J^n \, = \, \hat{Q} \, \chi_J^{n-1} &=&
     \left [ \hat{Q} \, , \, \gamma_0^{2J+1} e^{-\sqrt2 (2J+1) \varphi} \right]
     \, \Psi_{-J-1}^{n-1} \nonumber\\
&\sim&
      \eta_0 \,\gamma_0^{2J} \,e^{-\sqrt2 (2J+1) \varphi}
      \,\Psi_{-J-1}^{n-1} \nonumber\\
&\sim&
      \eta_0 \,(j^-)^{2J} e^{-\sqrt2 (2J+1) \varphi}
      \,\Psi_{-J-1}^{n-1} \stop
\eeqar
Note that this new state is at the same level as the original state,
and its ghost number is one larger because the $\ket{\da}_\gh$
is raised by $\eta_0$.  Also, the new state is a {\it LWS \/} of spin $J$,
since the HWS has been lowered $2J$ times.  Comparing graphs~2a and 2b, one
can see that these new states obtained by the Felder procedure are simply
the LWS that we already know.  The reason for this is that the HWS and LWS
are conjugate states in the same representation of the current algebra,
for $2J+1 \in \NN$, but are not in the same representation in the Fock space.
They can therefore be found from both sectors of the Fock space, after
performing the Felder resolution.

On the other hand, when $J \le -\thrhalf$ one gets genuine new states.
We have already seen this at the first level.  Particularly interesting states
of this type are the states of ghost number 3, with conjugates of ghost
number $-2$, since they
can not have any analogues in the conformal gauge.  As can be seen
from graph~2b, the first such state occurs at $J=-\fhalf$, and comes
from the $\Psi_{1/2}^2$ state
$c_{-1}\ket{\gamma_0=0,p_\varphi= -2 i\sqrt2, p=0} \upgh$ of
table~\ref{tab}.  In this case, the cosingular vector $\chi_{-5/2}^2$
satisfying $Q_3 \chi_{-5/2}^2 =\Psi_{1/2}^2$ can be taken to be:
\beq
\chi_{-5/2}^2 = c_{-1} \gamma_{-1}^3 \ket{\gamma_0=0,p_\varphi= i\sqrt2, p=0}
        \upgh \stop
\eeq
Then the desired new ghost number 3 state is
\beqar
\Psi_{-5/2}^3 &=& \hat{Q} \, \chi_{-5/2}^2 \nonumber\\
   &=& \eta_{-1} c_{-1} \gamma_{-1}^2
   \ket{\gamma_0=0, p_\varphi= i\sqrt2, p=0} \upgh \nonumber\\
   &\to& \eta_{-1} c_{-1} (J_{-1}^+)^2 \ket{J=-\fhalf,M=\fhalf, p=0} \upgh
\stop
\label{state}
\eeqar
Using the procedures given above, one can also obtain its
conjugate state with ghost number $-2$:
\beq
\Psi_{-5/2}^{-2} =
      \zeta_{-1} \left( b_{-1} J_{-1}^- + 2 \zeta_{-2} \right) J_{-1}^-
      \ket{J=-\fhalf,M=-\fhalf, p=0} \downgh \stop
\eeq
One can check that this state is not exact, and that it closes up to a
singular vector.

\heading{The absolute cohomology on the current algebra}

The only way to avoid the conclusion that the current algebra cohomology is
different than that of the Fock space, would be for these extra states
with funny ghost numbers to not appear in the absolute cohomology.
The absolute cohomology can be found from the relative
cohomology using a long exact sequence,
with the operators $i$ (inclusion),
$b_0$ and $M = \sum_{n\neq 0} n \, c_{-n} c_n$ \cite{seewit}.
If the relative cohomology exists at
only two consecutive ghost numbers the operator $M$ must be trivial, so
one ends up with a short exact sequence.  This means that
the absolute cohomology is simply a doubling of the relative
cohomology.  In our case the cohomology exists at several ghost
numbers, so the argument does not go through.  However, for the highest
ghost number states, the exact sequence reduces to
\beq
0 = H_{rel}^4 \onarrow{i} H_{abs}^4 \onarrow{b_0} H_{rel}^3
\onarrow{M} H_{rel}^5 = 0 \comma
\eeq
so $H_{abs}^4 \simeq H_{rel}^3$.  At
$J=-\fhalf$, the ghost number 4 state of the absolute cohomology is given
simply by raising the relative-cohomology state of eq\eq{state}
with $c_0$, giving:
\beq
\eta_{-1} c_{-1} (J_{-1}^+)^2 \ket{J=-\fhalf,M=\fhalf, p=0} \upgh
\otimes\ket{\uparrow}_{bc} \stop
\eeq
Since there are no states with ghost number larger than 2 in the absolute
cohomology of the conformal
gauge, the existence of this state proves that

\noindent
{\it The BRST cohomology in the light-cone gauge, with the gravity state
space being the irreducible $SL(2,R)$ Kac-Moody modules, is inequivalent to
that of the conformal gauge.\/}
\noindent

We conclude that one should not work on irreducible $SL(2,R)$
Kac-Moody representations in the gravitational sector.   It is interesting
that a similar conclusion also holds in the conformal gauge, where the
Liouville field is not taken to give an irreducible representation of the
Virasoro algebra.  It is also the case in topological $G/G$ theories; this
is implicit in the works of refs.~\cite{Gaweds,Spiegel}, and can be seen in
ref.~\cite{G/G}.

\section{Conclusions}

In order to study the BRST cohomology of a theory coupled to gravity, one
needs to define the state space of the gravity sector.  In the conformal
gauge in two dimensions the gravity sector is represented by the Liouville
field, which is usually quantized as a free scalar with a background charge. In
Polyakov's light-cone gauge the gravity sector is represented by an
$SL(2,R)$ Kac Moody algebra.  In this case there are two reasonable
choices for the gravity state space: one can work on irreducible $SL(2,R)$
modules, or on the Fock space of the (modified) Wakimoto free-field
representation of the $SL(2,R)$.  These two choices lead to different
cohomologies.  We have analyzed the cohomology of the $c=1$ theory on both
these spaces, and have shown how the cohomology on the current algebra can
be derived from that on the Fock space using a Felder resolution.

In the Wakimoto representation, the theory contains a scalar field
$\varphi$ playing the role of the Liouville field.  In addition, there
are extra fermionic ghost fields $(\eta,\zeta)$ and extra bosonic
fields $(\beta,\gamma)$.  Unlike the conformal gauge, the quantization
of all these fields is straightforward, since the measure in the
gravity sector is field independent.  Since
the theory in the Wakimoto representation
has more fields and a more complicated structure than the Liouville
theory, it is probably
impractical for amplitude calculations.  It
is possible, however, to find the BRST cohomology in this gauge,
and we see that {\it the spectrum of the light-cone theory on the
Fock space is equivalent to that of the Liouville theory.\/}
The operators generating the cohomology are
identical to those of the Liouville theory, except for the generators
of the ground ring.  The light-cone ground ring operators depend on
the extra fields,
but their algebra is identical to that of the conformal gauge.

The BRST cohomology on irreducible $SL(2,R)$ Kac-Moody modules leads to a
cohomology structure and spectrum different than that of the conformal
gauge.  In particular, the cohomology contains operators with
ghost numbers up to  $3$ in the relative cohomology, and up to $4$ in the
absolute cohomology.  These operators have no analogues in
the conformal gauge.  Since the Fock space cohomology agrees with that of
the conformal gauge, and one would like to obtain gauge-independent results,
we conclude that one should not work with irreducible representations in the
gravitational sector.

In order to describe $c<1$ matter theories coupled to 2d gravity, the
free scalar describing the matter of the $c=1$ theory
should be replaced by a coloumb
gas description.  The gravity sector is still built from the Wakimoto
fields representing the current algebra.  As in the conformal gauge,
the Fock-space cohomology consists only of tachyon states \cite{Pilch}.
After employing the Virasoro-algebra Felder resolution \cite{Felder}
for the matter sector, one will see that the cohomology in the Fock space is
the same as that of the conformal gauge \cite{Zuck,Pilch}.  In fact, one has
the stronger result that the same
operators can represent the
cohomology in both gauges.  As in the $c=1$ case,  one could now find the
cohomology on the irreducible $SL(2,R)$ modules using the Felder
resolution for the gravity sector, and the resulting cohomology would be
different from that of the conformal gauge.

In conclusion, we have seen that  the cohomology structure in the light-cone
gauge, with the appropriate choice of state space, is identical to that of the
conformal gauge.  This supports the claim that discrete states are physical
objects.

\vskip 1 cm

{\large \bf \noindent Acknowledgments}

We are grateful to O.~Aharony, J.~Sonnenschein, S.~Yankielowicz and N.~Sochen,
for many useful discussions.

\newpage

\appendix{First level BRST cohomology on the $SL(2,R)$ Kac-Moody module}

In this appendix we calculate explicitly the first level relative BRST
cohomology of $\hat{Q}$ on the $SL(2,R)$ Kac-Moody modules. The expressions
for the modes of the various stress tensors needed for the calculation are:
\beqar
L_n^{matt} &=& \f{1}{2}\sum_m \, \l: x_m x_{n-m} \r: \nonumber\\
L_n^{bc\phantom{av}} &=& \sum_m \, (2n-m) \l: b_m c_{n-m} \r: \nonumber\\
L_n^{\gh\phantom{av}} &=& -\sum_m \, m \l: \zeta_m \eta_{n-m} \r:
\nonumber\\
L_n^{grav} &=& \sum_m \, \l: \left( \half J^-_m J^+_{n-m}+
\half J^+_m J^-_{n-m} - J^0_m J^0_{n-m} \right) \r:
- (n+1) \, J_n^0 \stop
\eeqar
It is also useful to have the commutation relations of the Virasoro
and Kac-Moody operators (recalling, from eq\eq{not primary} that $J^0$
is not a primary field):
\beq
[L_n,J_m^a] = (n a -m)J^a_{n+m} -\f{\kappa}2 \, n (n+1)
\, \delta_{n+m,0} \, \delta_{a,0} \stop
\eeq

At the first level, the relative BRST operator $\hat{Q}$ reduces to:
\beqar
\hat{Q} &\sim& \eta_0 \left( J_0^- +c_1 \zeta_{-1} - c_{-1} \zeta_1 \right)
+ \eta_{-1} J_1^- + \eta_1 J_{-1}^- \nonumber\\
&& {} + c_{-1} \left( L_1^{grav} + p \, x_1 \right)
  + c_1 \left( L_{-1}^{grav} + p \, x_{-1} \right) \stop
\eeqar
Since the operators are conformally normal ordered, $L_{-1}^{grav}$ is given by
\beq
L_{-1}^{grav} \sim J_0^- J_{-1}^+ + J_{-1}^- J_0^+ - 2 J_{-1}^0 J_0^0 \stop
\label{L1}
\eeq
Note that the order of the $J$'s is not the naive one.
Also, at the first level, the $L_0=0$ condition of eq\eq{L0} means that
\beq
p=0 \iff J = \half ~{\rm or} ~ J = \thrhalf \stop \label{app}
\eeq
It will turn out that $p$ vanishes for all the states in the cohomology
at this level, and it will be suppressed in writing the states.

\heading{Ghost number --1}
\beqar
\hat{Q} (\zeta_{-1}\ket\da) &=& J_{-1}^-\ket\da - J_0^-\zeta_{-1}
\ket\ua\nonumber\\
\hat{Q} (b_{-1}\ket\da) &=&  ( p \, x_{-1}+L_{-1}^{grav} )
\ket\da + (\zeta_{-1} -J_0^- b_{-1} ) \ket\ua
\eeqar
For the state $\zeta_{-1}\ket\da$ to be closed, it must be a heighest
weight state (HWS) and must be annihilated by $J_{-1}^-$.  Upon acting
with $J_1^+$, the latter condition yields $M=-\thrhalf $, using the
commutation relations of eq\eq{comm} and the definition of $M$ from
eqs\eq{su2}.  Since $J=-\thrhalf$, $p=0$ by eq\eq{app}.  In addition,
using eq\eq{L1}, one can verify that the state $(b_{-1}-\zeta_{-1}
J_0^+)\ket{\half,\half}$ is closed. Since there are no states of ghost
number lower than $-1$, the closed states are not exact and thus belong
to the relative cohomology.

\heading{Ghost number 0}
\beqar
\hat{Q} (b_{-1}\ket\ua) &=& (p\, x_{-1}+ L_{-1}^{grav})\ket\ua\nonumber\\
\hat{Q} (x_{-1}\ket\da) &=& J_0^-x_{-1}\ket\ua + p\, c_{-1}
\ket\da
\eeqar
For the state $b_{-1}\ket\ua$ to be closed
$p$ must vanish, and $L_{-1}^{grav}$ on the state must give 0.  Using
eq\eq{L1}, we see that the state must be a lowest weight state (LWS)
with $J=-M=-\thrhalf $.  This state
corresponds to the identity operator. The state $x_{-1}\ket\da$ is
closed provided it is a HWS with $p=0$.  Thus, again, $J = -\thrhalf $
or $J = \half $. One can verify that all these states are not exact.

\heading{Ghost number 1}
\beqar
\hat{Q} (x_{-1}\ket\ua) &=& p\, c_{-1}\ket\ua\nonumber\\
\hat{Q} (c_{-1}\ket\da) &=& -J_0^-c_{-1}\ket\ua
\eeqar
The first state is closed for $p = 0$, and so has $J = -\thrhalf $ or
$J = \half $.  Unless it is a LWS, it is exact, being
$\hat{Q} (x_{-1}\ket\da)$.
The second state is closed if it is a HWS. It is
then exact, $\hat{Q} (x_{-1}\ket\da)$, unless $p = 0$.  It must have
$J=-\thrhalf $, since the  $J = \half $ state is exact, being $\hat{Q}
(J_{-1}^0\ket\da)$.

\heading{Ghost number 2}
\beqar
\hat{Q} (c_{-1}\ket\ua) &=& 0\nonumber\\
\hat{Q} (\eta_{-1}\ket\ua) &=& 0
\eeqar
These states are trivially closed.
$c_{-1}\ket\ua$ is exact, $\hat{Q} (c_{-1}\ket\da)$, unless
it is a LWS.  It is also exact, $\hat{Q} (x_{-1}\ket\ua)$, unless $p=0$.
Using:
\beqar
\hat{Q} (\eta_{-1}\ket\da) &=&
	c_{-1}\ket\ua - J_0^-\eta_{-1}\ket\ua \nonumber\\
\hat{Q} (J^+_{-1}\ket\ua) &=&
	-2(M-\thrhalf)\eta_{-1}\ket\ua + 2c_{-1}J_0^+\ket\ua \nonumber\\
\hat{Q} (J_{-1}^0\ket\ua) &=&
	( - J_0^-\eta_{-1}\ket\ua + (M+\thrhalf)c_{-1} \ket\ua \comma
\label{eta}
\eeqar
one sees that the $J=-\thrhalf$ state is exact, being $\hat{Q}
(\eta_{-1}\ket\da -J_{-1}^0\ket\ua)$.  Thus, $J=\half $.  Using
eqs\eq{eta}, one can also see that the state $\eta_{-1}\ket\ua$ is exact
unless it is a LWS with $J=-\thrhalf$.

There are no other nontrivial cohomology states at this level.

\appendix{A proof of theorem~7 via spectral sequences}

In this appendix we briefly introduce the concept of spectral
sequences in association with double complexes \cite{Bott} and use
it to prove theorem~7.

A spectral sequence is a sequence of two-dimensional arrays of abelian groups
$E_r = \{E_r^{p,q}; p,q\in \ZZ\} ~r=1,2,\ldots$, with group homomorphisms
$d_r$ that map the array to itself:
\beq
d_r : E_r^{p,q} \rightarrow E_r^{p+r,q-r+1} \stop
\eeq
A property of the spectral sequence is that $E_{r+1}$ is
the cohomology of $E_r$ with respect to the map $d_r$.
Moreover, it has a well defined limit $E^{pq}_\infty=\lim_{r\rightarrow\infty}
E_r^{p,q}$.

A double complex is a complex with two anticommuting operators  $d$ and
$\delta$ that raise two gradings $q$ and $p$, respectively.   One can also
define the operator $D=d+\delta$, satisfying $D^2=0$, that  raises the total
grading $n=p+q$. To any double complex one can associate a spectral sequence
\beqar
E_1^{p,q} \, &\equiv& \, H_d^{p,q}\nonumber\\
E_2^{p,q}\, &\equiv& \, H_{\delta}^p(H_d^q)\nonumber\\
&\vdots& \comma
\eeqar
which converges to the cohomology of $D$:
\beq
H^n_D = \oplus_{p+q=n}E_{\infty}^{p,q}\stop
\label{D}
\eeq

One can relate $H_{\delta}(H_d)$ to $H_d(H_\delta)$ by using the spectral
sequence first to calculate the cohomology of the double complex directly,
and then after interchanging the roles of $d$ and $\delta$. Thus,

\noindent{\bf Theorem:} { \it
If the Felder complex exists only at degrees 0 and either $+1$ or $-1$, then
$H_{rel}^n (H_F^0) = H_F^0 (H_{rel}^n) + H_F^{\pm 1} ( H_{rel}^{n \mp 1} )$.
}

\noindent
{\bf Proof}:  In the definition above, let $d$ be the Felder operator
and  $\delta$ the relative BRST operator: $d=Q_F, \delta = \hat{Q}$.
Then $E_1^{p,q} = H_F^{p,q}$ vanishes for all $q \ne 0$, since the Felder
cohomology exists only at Felder index 0.  This means that
$E_2^{p,q} = H_{rel}^p (H_F^q)$ also is nontrivial only for $q=0$.
Now, $E_3^{p,q}$ is defined as the the cohomology of
$E_2^{pq}$ with respect to the operator $d_2:E_2^{p,q}
\rightarrow E_2^{p+2,q-1}$.  Since this changes the value of $q$, the
operator must vanish and $E_3^{pq} = E_2^{pq}$.  Therefore the sequence has
converged:
\beq
E_r^{pq} = E_2^{pq} = H_{rel}^p (H_F^0) \delta_{q,0}
		\quad \forall r\geq 2 \stop
\label{En}
\eeq
Using eqs\eq{D} and (\ref{En}), the cohomology of the double complex is
given by:
\beq
H_D^n = H_{rel}^n(H_F^0) \stop
\label{x17}
\eeq

Now let $d=\hat{Q}, \delta = Q_F$.  In this case,
\beq
E_2^{pq} = H_F^p(H_{rel}^q)
\label{En2}
\eeq
vanishes for $p\neq 0,\pm1$, since the
Felder complex exists only at $p=0$ and either $p=+1$ or $p=-1$.
Since $d_2:E_2^{p,q}\rightarrow E_2^{p+2,q-1}$ shifts the Felder index by 2,
it again vanishes, so the sequence has converged.  Eqs\eq{D} and (\ref{En2})
now show that
\beq
H_D^n = H_F^0(H_{rel}^n) + H_F^{\pm1}(H_{rel}^{n\mp1}) \stop\label{D1}
\eeq
The theorem is proved by equating
eqs\eq{x17} and (\ref{D1}).

\end{document}